\documentclass[twocolumn,nofootinbib]{openjournal}

\usepackage{graphicx}	
\usepackage{amsmath}	
\usepackage{amssymb}	

\usepackage{bm}         
\usepackage[ruled,vlined]{algorithm2e}
\usepackage{environ}
\usepackage{newtxtext,newtxmath}
\usepackage{natbib}
\usepackage[dvipsnames]{xcolor}
\usepackage{bibentry}
\makeatletter\let\saved@bibitem\@bibitem\makeatother
\usepackage[pdftex]{hyperref}
\hypersetup{final=true,
colorlinks=true,
linkcolor=Mahogany,
citecolor=Periwinkle}
\usepackage{savesym}
\savesymbol{tablenum}
\usepackage{siunitx}
\restoresymbol{SIX}{tablenum}
\usepackage{tikz}
\usepackage{tkz-euclide}

\usetikzlibrary{arrows.meta}

\newcommand{\posterior}[1][]{\ensuremath{p(\kappa|\gamma)}}
\newcommand{\likelihood}[1][]{\ensuremath{p(\gamma|\kappa)}}

\makeatletter
\newsavebox{\measure@tikzpicture}
\NewEnviron{scaletikzpicturetowidth}[1]{%
  \def\tikz@width{#1}%
  \def\tikzscale{1}\begin{lrbox}{\measure@tikzpicture}%
  \BODY
  \end{lrbox}%
  \pgfmathparse{#1/\wd\measure@tikzpicture}%
  \edef\tikzscale{\pgfmathresult}%
  \BODY
}
\makeatother

\begin{document}
\title[Trans-dimensional mass-mapping]{Sparse Bayesian mass-mapping using trans-dimensional MCMC}

\author{
Augustin Marignier,$^{1,2,3\dagger}$
Thomas Kitching,$^{2}$
Jason D. McEwen$^{2}$
Ana M. G. Ferreira$^{3,4}$}

\email{$\dagger$ Correspondence address: augustin.marignier@anu.edu.uk}

\affiliation{$^{1}$Research School of Earth Sciences, Australian National University, Canberra 2601, Australia\\$^{2}$Mullard Space Science Laboratory, University College London, Dorking RH5 6NT, UK\\
$^{3}$Department of Earth Sciences, University College London, London WC1E 6BT, UK \\
$^{4}$ CERIS, Instituto Superior T\'{e}cnico, Univesidade de Lisboa, Lisbon, Porturgal}

\begin{abstract}
    Uncertainty quantification is a crucial step of cosmological mass-mapping that is often ignored.
    Suggested methods are typically only approximate or make strong assumptions of Gaussianity of the shear field.
    Probabilistic sampling methods, such as Markov chain Monte Carlo (MCMC), draw samples form a probability distribution, allowing for full and flexible uncertainty quantification, however these methods are notoriously slow and struggle in the high-dimensional parameter spaces of imaging problems.
    In this work we use, for the first time, a trans-dimensional MCMC sampler for mass-mapping, promoting sparsity in a wavelet basis.
    This sampler gradually grows the parameter space as required by the data, exploiting the extremely sparse nature of mass maps in wavelet space.
    The wavelet coefficients are arranged in a tree-like structure, which adds finer scale detail as the parameter space grows.
    We demonstrate the trans-dimensional sampler on galaxy cluster-scale images where the planar modelling approximation is valid.
    In high-resolution experiments, this method produces naturally parsimonious solutions, requiring less than 1\% of the potential maximum number of wavelet coefficients and still producing a good fit to the observed data.
    In the presence of noisy data, trans-dimensional MCMC produces a better reconstruction of mass-maps than the standard smoothed Kaiser-Squires method, with the addition that uncertainties are fully quantified.
    This opens up the possibility for new mass maps and inferences about the nature of dark matter using the new high-resolution data from upcoming weak lensing surveys such as \emph{Euclid}.
\end{abstract}

\maketitle



\section{Introduction}
\label{sec:intro}
Gravitational lensing is the phenomenon where light from distant objects is distorted by the density field between the object and the observer, resulting in stunning images of warped, sheared, magnified and multiplied galaxies.
Lensing is due to both light and dark forms of matter, making it a promising probe of the nature of dark matter \citep{Heavens2009}.
Weak lensing is the regime of small distortions, the effect of which on distant galaxies can be described as a magnification due to a convergence field \(\kappa\) and a perturbation of intrinsic ellipticity due to a shear field \(\gamma\).

The convergence field \(\kappa\) is the integrated total mass density along the line of sight \citep{Bartelmann2001,Dodelson2017}, and hence is a measure of the matter over-density field.
The density field contains both Gaussian and non-Gaussian structures.
The Gaussian structures of the cosmological initial conditions evolve under the non-linear influence of gravity to produce non-Gaussian structures.
As such, it is essential that methods to build maps of the convergence field (mass maps) contain Gaussian and non-Gaussian structures.
With non-Gaussian structures, the higher-order statistics such as Minkowski functions and bispectrum \citep{Munshi2017} of the mass maps can be measured in addition to the usual two-point correlation function \citep{Peebles1980}.
From these statistics, one can compare with the predictions of the statistical density distributions of $\Lambda$CDM or other cosmologies.
Further, one can compare the total density distribution in colliding cluster systems with the observed distribution of baryonic matter to infer the presence and collisional properties of dark matter \citep[e.g.][]{Clowe2006,Harvey2015}.
These sorts of comparisons would be much better informed if uncertainties were also quantified with the mass maps.

Unfortunately, \(\kappa\) cannot be directly observed, and must be constrained by an ill-posed inverse problem of observed ellipticities modelling the shear field \(\gamma\).
There have been many methods proposed to solve this inverse problem.
The standard method is the Kaiser-Squires \citep[KS,][]{Kaiser1993}, which is a simple Fourier space linear operator.
However, it is well known that this method is not robust to noise and missing data, and is often smoothed to mitigate the effects of noise at the expense of small-scale non-Gaussian structures.
Despite this the mass maps  recovered with KS are generally reliable, and the method has been extended to the sphere \citep{Wallis2021} for use in wide-field surveys \citep{VanWaerbeke2013,Chang2018,Jeffrey2018,Jeffrey2021,Price2020b}.
Recently, there has been a push for methods with sparsity-promoting regularisation to preserve non-Gaussianities and deal with irregular data sampling \citep[e.g.][]{Lanusse2016,Price2021,Starck2021}, where sparsity is typically promoted in a wavelet basis.
Sparsity is the property where many of the coefficients representing a signal in a given basis are (close to) zero --- a property empirically observed in many images \citep[see e.g.][]{Donoho2006,Candes2011}.

Despite all these methods, uncertainty quantification for mass-mapping remains a difficult task.
Common approaches for uncertainty quantification involve computationally expensive posterior sampling by, for example, Markov chain Monte Carlo (MCMC) methods.
In the weak lensing context, however, only recently have these methods been used for mass-mapping \citep[e.g.][]{Alsing2016,Price2020,Porqueres2021,Fiedorowicz2022,Remy2022}.
This is explained by imaging inverse problems being very high-dimensional, resulting in difficulties in converging to the posterior.
To make the sampling of the posterior computationally tractable, the slope of the posterior is often exploited to identify the high-density regions, for example in Hamiltonian Monte Carlo \citep{Neal2012} or proximal MCMC \citep{Pereyra2016}.
Both of these approaches have now been used for mass-mapping \citep{Price2020,Fiedorowicz2022,Remy2022,Marignier2023}.
Various priors have been used, including Gaussian \citep{Alsing2016}, sparsity \citep{Price2020,Marignier2023}, physically informed power spectrum \citep{Porqueres2021}, log-normal \citep{Fiedorowicz2022} and learnt \citep{Remy2022} priors.

The purpose of this work is to introduce a new efficient MCMC sampling method for mass-mapping.
We use a trans-dimensional MCMC sampler to gradually grow the parameter space as required by the data.
Trans-dimensional inversions, where the number of model parameters it itself a parameter to be determined \citep{Geyer1994,Green1995}, are now extremely common in geophysical studies \citep[e.g.][]{Malinverno2002,Gallagher2012,Bodin2009,Minsley2011,Tkalcic2013,Piana2015,Hawkins2015,Burdick2019}, have been used sporadically in astronomy \citep[e.g][]{Cornish2007,Karnesis2014,Feder2018} and have yet to be used for mass-mapping.
The particular sampler we use was proposed by \citet{Hawkins2015} for geophysical imaging, and uses a wavelet tree parameterisation allowing for multi-scale analysis and sparsity-promotion.
The appeal of this method is its ability to significantly reduce the effective parameter space.
For example, \citet{Hawkins2015} showed a 3D seismic tomography example with a parameter space of at most 524,288 parameters, and their method produced a solution with around 500 parameters.
This sort of efficiency in sampling the parameter space will be essential for high-resolution mass-maps as the sky-fraction covered by upcoming surveys increases.
Other methods for sampling such high parameter spaces typically require gradient or proximal mapping calculations \citep[e.g][]{Neal2012,Pereyra2017}, which are avoided with this trans-dimensional method.

In this work we use the trans-dimensional tree MCMC sampler of \citet{Hawkins2015} for Bayesian mass-mapping with full uncertainty quantification.
We exploit the multi-scale nature of the wavelet parameterisation to impose a scale-dependent sparsity-promoting prior using a generalised Gaussian distribution (GGD), as opposed to the more common Laplace distribution.
The rationale for this is that to first order larger scale structures in mass maps are more Gaussian than the smaller scale details.
A similar rationale was used by \citet{Starck2021} for mass mapping, although our approach is more similar to that used by \citet{McEwen2017} for the analysis of cosmic strings in the cosmic microwave background.
We demonstrate this method on the recovery of cluster-scale convergence maps from N-body simulations.

This article is organised as follows.  In \autoref{sec:background} we outline the theoretical background of weak lensing and Bayesian inversion.  In \autoref{sec:tdt} we give details of the trans-dimensional tree inversion.  \autoref{sec:tdtmm} outlines the specificities of the mass-mapping inverse problem.  \autoref{sec:sims} presents our results on simulation data, which we discuss further in \autoref{sec:discussion}.  We conclude in \autoref{sec:conclusions}.

\section{Background}
\label{sec:background}
\subsection{Weak Lensing and Mass-Mapping}
Gravitational lensing describes the deflection of photons from distant sources as they pass by regions of local gravitational potential variations caused by the local over or under density of matter acting as a lens.
The weak lensing effect can be described in terms of a 3D lensing potential \(\psi(\bm{r}) = \psi(\theta_1, \theta_2, r)\), which is the integral along the line of sight of the Newtonian gravitational potential \(\Phi(\bm{r})\)
\begin{equation}
    \psi(\bm{r}) = \frac{2}{c^2}\int_0^r \frac{f_K(r - r')}{f_K(r)f_K(r')}\Phi(r'\theta_1, r'\theta_2, r)dr'.
\end{equation}
Here, \(r\) is the comoving distance from the observer to the source, \((\theta_1, \theta_2)\) are angular positions on the sky, \(c\) is the speed of light in a vacuum, and \(f_K(r)\) is the comoving angular separation in a cosmology with curvature \(K\), which we take to be 0 (i.e.\ a flat universe).
The gravitational potential is related to the fractional matter over-density field \(\delta(\bm{r})\) by Poisson's equation
\begin{equation}
    \nabla^2\Phi = \frac{3\Omega_m H_0^2}{2c^2a(t)}\delta(\bm{r})
\end{equation}
where \(\Omega_m\) is the current matter density parameter, \(H_0\) is the current Hubble constant, and \(a(t)\) is the scale factor.
The aim of mass-mapping is to infer the convergence field \(\kappa(\bm{\theta})\) from galaxy distortion measurements representing the complex shear field \(\gamma(\bm{\theta}) = \gamma_1(\bm{\theta}) + i\gamma_2(\bm{\theta})\).
The convergence represents the total density perturbation along the line of sight.
The convergence and the shear are defined as gradients of the lensing potential \(\psi(\bm{r})\)
\begin{align}
    \kappa &\equiv \frac{1}{2}\left(\frac{\partial^2\psi}{\partial\theta_1^2} +  \frac{\partial^2\psi}{\partial\theta_2^2}\right) \nonumber\\
    \gamma_1 &\equiv \frac{1}{2}\left(\frac{\partial^2\psi}{\partial\theta_1^2} -  \frac{\partial^2\psi}{\partial\theta_2^2}\right) \nonumber\\
    \gamma_2 &\equiv \frac{1}{2}\left(\frac{\partial^2\psi}{\partial\theta_1\theta_2}\right).
    \label{eqn:kappagammaPhi}
\end{align}
Evaluating the derivative in Fourier space and rearranging gives a linear relation between the shear and convergence 
\begin{equation}
    \label{eqn:KS_tdt}
    \tilde{\gamma}(\bm{l}) = \frac{l_1^2 - l_2^2 + 2i l_1 l_2}{l_1^2 + l_2^2}\tilde{\kappa}(\bm{l}),
\end{equation}
where \(\bm{l} = (l_1,l_2)\) is the conjugate variable of \(\bm{\theta} = (\theta_1, \theta_2)\).
This defines the Fourier space KS kernel \citep{Kaiser1993}.
Note that it is undefined at the origin \(l_1 = l_2 = 0\), which corresponds to the mass-sheet degeneracy.
The mass-sheet degeneracy implies that \(\kappa\) can only be determined up to an additive constant.
Physically, this is because a constant surface mass density does not cause any shear \citep{Bartelmann2001}.

\subsection{Bayesian Mass-mapping}
From Bayes' Theorem, the \textit{posterior} probability distribution describing the probability that the underlying convergence field producing observed complex shear measurements \(\gamma\) is described by some discretised field \(\kappa\), is given by
\begin{equation}
    p(\kappa|\gamma) \propto p(\gamma|\kappa)p(\kappa)
\end{equation}
where \(p(\gamma|\kappa)\) is the likelihood function describing data fidelity between the observed \(\gamma\) and predictions generated from \(\kappa\), and \(p(\kappa)\) is a distribution describing any prior belief about the convergence field, e.g.\ that it is sparse in a wavelet basis (see \hyperref[subsec:prior]{subsections~\ref{subsec:prior}}~and~\ref{subsec:ggd}).
The pixels of the \(\kappa\) map, or their representation in a particular basis, are parameters to be inferred.

In standard MCMC, a sampler will start at some model of convergence \(\kappa_i\) and propose some new model \(\kappa_i'\) based on a symmetric proposal distribution \(q(\kappa_i'|\kappa_i)\).
This distribution is symmetric in that \(q(\kappa_i'|\kappa_i) = q(\kappa_i|\kappa_i')\), i.e.\ the probability of proposing \(\kappa_i'\) when starting at \(\kappa_i\) is the same as the probability of proposing \(\kappa_i\) when starting at \(\kappa_i'\).
The proposed model is either accepted (\(\kappa_{i+1} = \kappa_i'\)) or rejected (\(\kappa_{i+1} = \kappa_i\)) according to an acceptance criterion.
A common choice of acceptance criterion is known as the Metropolis-Hastings (MH) criterion
\begin{equation}
    \alpha(\kappa',\kappa) = \min\left\{1, \frac{p(\kappa')p(\gamma'|\kappa')q(\kappa_i|\kappa_i')}{p(\kappa_i)p(\gamma|\kappa_i)q(\kappa_i'|\kappa_i)}\right\}
\end{equation}
Repeating this many thousands of times produces a chain of samples which, thanks to the acceptance criterion, is guaranteed to converge to the posterior distribution, giving the solution of the ill-posed inverse problem of inferring the convergence from the shear.
A summary convergence map (e.g.\ the sample mean) can be calculated, as can any measure of uncertainty.
This is a great strength of sampling methods, such as MCMC, although it comes at great computational cost.
We note here that while this method involves predominantly forward modelling, our analysis is solving an inverse problem and hence will be referred to throughout as an inversion.

\subsection{Wavelet Transform}
\label{subsec:wavelet_transform}
The wavelet transform is a popular tool in image processing, having been popularised in the late 20\textsuperscript{th} century \citep{Mallat1989,Daubechies1992}.
Since then, the theory of compressed sensing \citep{Donoho2006,Candes2011} showed that sparse signals can be accurately recovered from incomplete data, leading many studies in different fields using sparsity-promoting priors or regularisation in a wavelet space, including weak lensing studies \citep[e.g.][]{Lanusse2016,Price2021,Starck2021}.

Wavelets are rapidly-decaying, wave-like functions that exist for a finite amount of time or space and form a basis in which typical signals tend to be sparse.
There are many different families of wavelets, the details of which are beyond the scope of this work.
The discrete wavelet transform can be seen as high and low-pass convolutions using scaled and translated versions of the ``mother'' wavelet, resulting in a separation of localised information at different length scales.
Typically these high and low pass filters extract the high and low half, respectively, of the frequencies in the image.
The outputs of the high-pass filters are the detailed wavelet coefficients, and the low-pass filters give the approximation coefficients.
Repeating the process on the approximation coefficients returns detailed wavelet coefficients at a slightly larger scale than the previous pass.
After each pass the wavelet coefficients are downsampled by a factor of two, as only half the frequency content from the previous scale remains.
The final set of approximation coefficients are known as the coefficients of the scaling function.
The scaling function is distinct from the wavelets as it accounts for the lowest frequency content, in particular at 0 frequency.

The remainder of this section gives a more mathematical description of the 2D discrete wavelet transform, largely summarising the seminal work of \citet{Mallat1989} in which further details can be found by the interested reader.

Beginning in one dimension, consider a function \(f(x)\in L^2(\mathbb{R})\).
Consider also a continuously differentiable and exponentially decreasing scaling function \(\phi(x)\), whose Fourier transform has the shape of a low-pass filter.
The projection of \(f\) onto the set of dilations and translations of \(\phi\) by a dyadic scale \(2^j\) for \(j\in\mathbb{Z}\) gives the \textit{approximation} of \(f\) at scale \(2^j\)
\begin{eqnarray}
    A_{2^j}f &=& (\langle f, \sqrt{2^j}\phi(2^jx - n)\rangle)_{n\in\mathbb{Z}} \nonumber\\
             &=& \left(\int_{-\infty}^{\infty}f(x)\sqrt{2^j}\phi(2^jx - n)dx\right)_{n\in\mathbb{Z}},
\end{eqnarray}
where the angled brackets denote the inner product on \(\mathbb{R}\).
This can be seen as the convolution of \(f\) with \(\phi\) evaluated at a spacing of \(2^{-j}\).
The \textit{detail} signal of \(f\) at scale \(2^j\) is the difference between approximations at scale \(2^j\) and the smaller scale \(2^{j+1}\).
It can be shown \citep{Mallat1989} that the detail can again be found as the convolution of \(f\) with some scaled and translated basis functions at a spacing of \(2^{-j}\).
For the detail, the basis is a wavelet function \(\xi(x)\), whose Fourier transform is a bandpass filter.
\begin{eqnarray}
    D_{2^j}f &=& (\langle f, \sqrt{2^j}\xi(2^jx - n)\rangle)_{n\in\mathbb{Z}} \nonumber\\
             &=& \left(\int_{-\infty}^{\infty}f(x)\sqrt{2^j}\xi(2^jx - n)dx\right)_{n\in\mathbb{Z}}.
\end{eqnarray}
The set of coefficients \(\{A_1f, D_{2^j}f\; \mathrm{for}\; 0 < j \leq J\}\) is the representation of \(f\) in the orthogonal wavelet basis.
In the discrete case, \(J=\log_2(N)\), where \(N\) is the length of \(f\).

In two dimensions, we consider scaling and wavelet functions that are separable in \(x\) and \(y\).
The 2D scaling function is given by \(\phi(x,y) = \phi(x)\phi(y)\), allowing for the extraction of both horizontal and vertical approximations.
This gives three wavelet functions for the horizontal, vertical and corner details
\begin{eqnarray}
    \xi^1(x,y) &=& \phi(x)\xi(y), \nonumber\\
    \xi^2(x,y) &=& \phi(y)\xi(x), \nonumber\\
    \xi^3(x,y) &=& \xi(x)\xi(y).
\end{eqnarray}
The approximation and details of \(f(x,y)\) are again obtained by projecting onto the sets of dilated and translated scaling and wavelet functions, giving the representation of \(f\) in the 2D orthonormal wavelet basis as \(\{A_1f, D^1_{2^j}f, D^2_{2^j}f, D^3_{2^j}f\; \mathrm{for}\; 0 < j \leq J\}\).

We omit here the details on the construction of specific scaling and wavelet functions and computing the transform in practice, as they are unnecessary to understand the mass-mapping method introduced in this work, and form an extensive literature in their own right.
We refer the interested reader to texts such as \citet{Mallat1989} and \citet{Daubechies1992} for these details.

\section{Trans-dimensional Tree Inversion}
\label{sec:tdt}
In trans-dimensional MCMC, the parameterisation and the dimensionality of \(\kappa\) is allowed to change \citep{Green1995}.
For example, the number of wavelet coefficients describing \(\kappa\) is variable.
This allows the available data to determine the complexity of the solution, rather than having this set \textit{a priori}.
Instinctively, it might seem that such a trans-dimensional inversion will naturally tend to more complex models, as one can obtain arbitrarily better fits to data by adding more and more model parameters.
It has been shown however that results are generally parsimonious.
To generalise the MH acceptance criterion, the determinant of a Jacobian matrix describing the transformation from one parameter space to another is included, i.e
\begin{equation}
    \label{eqn:alpha}
    \alpha(\kappa',\kappa) = \min\left\{1, \frac{p(\kappa_i')p(\gamma'|\kappa_i')q(\kappa_i|\kappa_i')}{p(\kappa_i)p(\gamma|\kappa_i)q(\kappa_i'|\kappa_i)}|\mathcal{J}|\right\}.
\end{equation}
Significantly though, as we outline in the following section, with simple choices about the parameterisation and proposal distribution the determinant of the Jacobian is 1, effectively keeping the original acceptance criterion.

The trans-dimensional sampler we use here is also known as a birth/death sampler, where model parameters are gradually added or removed, as well have their values perturbed as in standard MCMC \citep{Green1995}.
In this work, we use a parameterisation first proposed by \citet{Hawkins2015} for geophysical inversions.
Described as a tree structure, nodes can be added and removed from the tree via the birth and death proposals, respectively.
The nodes are arranged such that birth proposals add more detail to the model.
We give some details of the structure and the prior and proposal distributions here, although for brevity we only describe the choices made for our mass-mapping application.
For more details, we refer the reader to \citet{Hawkins2015}.
The aim of using this tree structure in the mass-mapping context is to gradually add more and more small-scale detail without having to sample an exceedingly large parameter space.

\subsection{The wavelet tree model}
\begin{figure}
    \centering
    \begin{scaletikzpicturetowidth}{\columnwidth}
        \begin{tikzpicture}[scale=\tikzscale,
                            squarednode/.style={rectangle,draw, minimum size=7mm},
                            treebranch/.style={-Stealth,color=red,line width=1mm},
                            treenode/.style={circle,minimum size=3mm,inner sep=0pt,fill=black}]

            \pgfmathsetmacro{\gridLength}{8}

            \node [squarednode,fill=blue!60] at (0.5,7.5) {};

            \node [squarednode,fill=orange] at (0.5,6.5) {};
            \node [squarednode,fill=orange!60] at (1.5,6.5) {};
            \node [squarednode,fill=orange!30] at (1.5,7.5) {};

            \foreach \y in {4,5}{
                \foreach \x in {0,1}
                    \node [squarednode,fill=black!40!green] at (\x+0.5,\y+0.5) {};
                \foreach \x in {2,3}
                    \node [squarednode,fill=black!40!green!60] at (\x+0.5,\y+0.5) {};
                }
            \foreach \x in {2,3}
                \foreach \y in {6,7}
                    \node [squarednode,fill=black!40!green!30] at (\x+0.5,\y+0.5) {};

            \foreach \y in {0,...,3}{
                \foreach \x in {0,...,3}
                    \node [squarednode,fill=yellow] at (\x+0.5,\y+.5) {};
                \foreach \x in {4,...,7}
                    \node [squarednode,fill=yellow!60] at (\x+0.5,\y+.5) {};
            }
            \foreach \x in {4,...,7}
                \foreach \y in {4,...,7}
                    \node [squarednode,fill=yellow!30] at (\x+0.5,\y+.5) {};

            \draw[thick] (1,8) -- (1,7) -- (0,7);
            \draw[thick] (2,8) -- (2,6) -- (0,6);
            \draw[thick] (4,8) -- (4,4) -- (0,4);
            \draw[thick] (8,8) -- (8,0) -- (0,0);

            \draw[dashed,thick] (1,6) -- (1,4);
            \draw[dashed,thick] (2,6) -- (2,0);
            \draw[dashed,thick] (3,8) -- (3,4);
            \draw[dashed,thick] (4,4) -- (4,0);
            \draw[dashed,thick] (0,2) -- (8,2);
            \draw[dashed,thick] (6,0) -- (6,8);
            \draw[dashed,thick] (2,6) -- (8,6);
            \draw[dashed,thick] (4,4) -- (8,4);
            \draw[dashed,thick] (0,5) -- (4,5);
            \draw[dashed,thick] (2,7) -- (4,7);

            \coordinate[treenode] (t0) at (0.5,7.5);
            \coordinate[treenode] (t1) at (0.5,6.5);
            \coordinate[treenode] (t2) at (1.5,6.5);
            \coordinate[treenode] (t3) at (1.5,7.5);
            \coordinate[treenode] (t4) at (2.5,7.5);
            \coordinate[treenode] (t5) at (3.5,6.5);
            \coordinate[treenode] (t6) at (5.5,6.5);
            \coordinate[treenode] (t7) at (3.5,5.5);
            \coordinate[treenode] (t8) at (1.5,4.5);
            \coordinate[treenode] (t9) at (3.5,1.5);
            \coordinate[treenode] (t10) at (2.5,0.5);

            \draw[treebranch] (t0) -- (t1);
            \draw[treebranch] (t0) -- (t2);
            \draw[treebranch] (t0) -- (t3);
            \draw[treebranch] (t3) -- (t4);
            \draw[treebranch] (t3) -- (t5);
            \draw[treebranch] (t4) -- (t6);
            \draw[treebranch] (t2) -- (t7);
            \draw[treebranch] (t1) -- (t8);
            \draw[treebranch] (t8) -- (t9);
            \draw[treebranch] (t8) -- (t10);

            \node[left] () at (0,7.5) {\(j=0\)};
            \node[left] () at (0,6.5) {\(j=1\)};
            \node[left] () at (0,5) {\(j=2\)};
            \node[left] () at (0,2) {\(j=3\)};

        \end{tikzpicture}
    \end{scaletikzpicturetowidth}
    \caption[Wavelet tree parameterisation]{The 2D wavelet tree parameterisation for a tree of maximum depth 3.
    Each coloured square corresponds to a wavelet coefficient, although only those with black dots form part of the current tree (red arrows).
    This current tree has \(k=11\) coefficients.
    The coefficients of each wavelet scale (\(j=\{0,1,2,3\}\)) are separated by solid black lines and grouped by colour (blues, oranges, greens, yellows).
    The blue coefficient at the root of the tree is the scaling coefficient.
    Beneath this we have 3 wavelet coefficients at scale \(j=1\).
    Beneath each of these three coefficients we have 4 coefficients at scale \(j=2\), where the relation is represented by the colour shade e.g.\ the 4 dark green coefficients in \(j=2\) are beneath the dark orange coefficient in \(j=1\).
    From each of the \(j=2\) coefficients we then get another 4 coefficients at scale \(j=3\), where the relation is now represented by dashed black lines e.g.\ the bottom-right set of 4 dark yellow \(j=3\) coefficients are directly below the bottom-right dark green \(j=2\) coefficient.
    Notice how the tree (red arrows) need not extend all the way down to the maximum depth.}
    \label{fig:wavtree}
\end{figure}
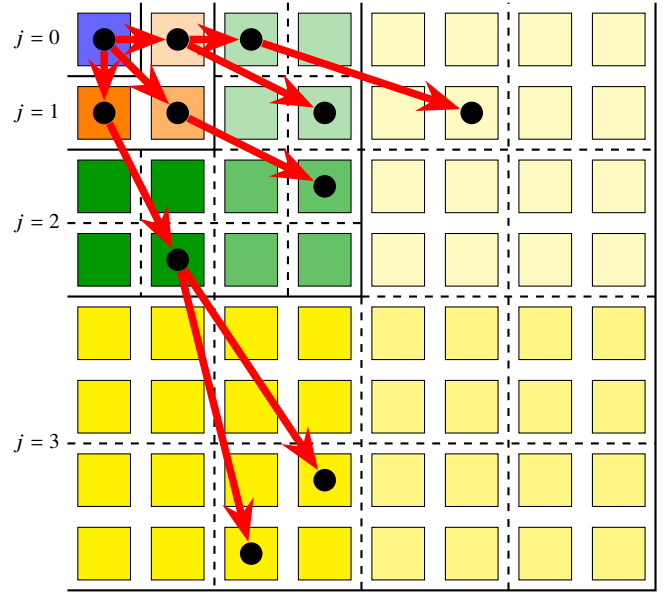

We parameterise our convergence maps using a set of \(k\) wavelet coefficients \(\varkappa\) in a tree structure, denoted by \(\mathcal{T}_k\).
The root of the tree is the scaling function coefficient, a single-pixel representation of the image.
The tree then branches down to smaller scale structure.
This is shown schematically in \autoref{fig:wavtree}.
The \(k\) coefficients of the current tree are indicated by black dots and their connection in the tree is indicated by the red arrows.
The maximum depth of the tree is the \(J\)\textsuperscript{th} wavelet scale, chosen based on the desired output resolution.
Our model space vector \(\kappa\) is then given by
\begin{equation}
    \kappa = \langle\mathcal{T}_k, \{\varkappa_i : i \in 1,\dots,k\}\rangle,
\end{equation}
where \(\langle\cdot,\dots,\cdot\rangle\) denotes a vector and \(i\) is a unique index for all the coefficients in the tree.

\subsection{The prior}
\label{subsec:prior}
The prior \(p(\kappa)\) consists of three parts: (1) the probability of the number of coefficients, \(p(k)\); (2) the probability that \(k\) coefficients are arranged in the tree \(\mathcal{T}_k\), \(p(\mathcal{T}_k|k)\); and (3) the prior on the wavelet coefficient values \(p(\varkappa_i|\mathcal{T}_k,k)\).

For the probability of \(k\) we choose a simple uniform prior
\begin{equation}
    p(k) = \frac{1}{k_{\max}}
\end{equation}
where we take \(k_{\max} = 2^J\), i.e.\ the theoretical number of wavelet coefficients with \(J\) wavelet scales.
This prior is effectively uninformative, as we have no initial prior belief as to how many wavelet coefficients should be in the tree.
This choice also simplifies the calculation of the proposal probabilities, which we describe in detail in \autoref{subsec:proposals}.

Similarly, we use a uniform prior for the arrangement of \(\mathcal{T}_k\).
Our tree is \textit{restricted and heterogeneous}.
Restricted trees have a maximum depth, in our case the number of wavelet scales \(J\).
Heterogeneous trees vary the number of child coefficients that each parent coefficient has.
For the 2D wavelet tree (\autoref{fig:wavtree}), the scaling coefficient (\(j=0\)) has 3 children (1 each for the horizontal, vertical and diagonal detail at the next scale), and then each coefficient from \(j=1\) onwards has 4 children, as the wavelet transform downsamples by a factor of 2 in both the vertical and horizontal directions (\autoref{subsec:wavelet_transform}).
Our prior is then
\begin{equation}
    p(\mathcal{T}_k|k) = \frac{1}{\mathcal{N}_{k,J}},
\end{equation}
where \(\mathcal{N}_{k,J}\) is the number of valid heterogenous tree configurations of \(k\) coefficients restricted to depth \(J\).
This is calculated via a recurrence relation, the details of which we omit here in the interest of space.
We refer the interested reader to \citet{Hawkins2015} for details, including a fast algorithm for computing prior ratios of tree structures.

We leave discussion of our value prior \(p(\varkappa_i|\mathcal{T}_k,k)\) to \autoref{subsec:ggd}, as this is more physically motivated rather than imposed by our parameterisation.
Suffice to say here that we use a generalised Gaussian prior on our wavelet coefficient values, with larger wavelets scales (low \(j\)) having a more Gaussian prior than smaller scales (high \(j\)).

These three priors are independent, so our overall prior \(p(\kappa)\) is given by the simple product
\begin{equation}
    p(\kappa) = p(\mathcal{T}_k|k)p(k)\prod_{i=1}^{k}p(\varkappa_i|\mathcal{T}_k,k).
\end{equation}

\subsection{Proposals}
\label{subsec:proposals}
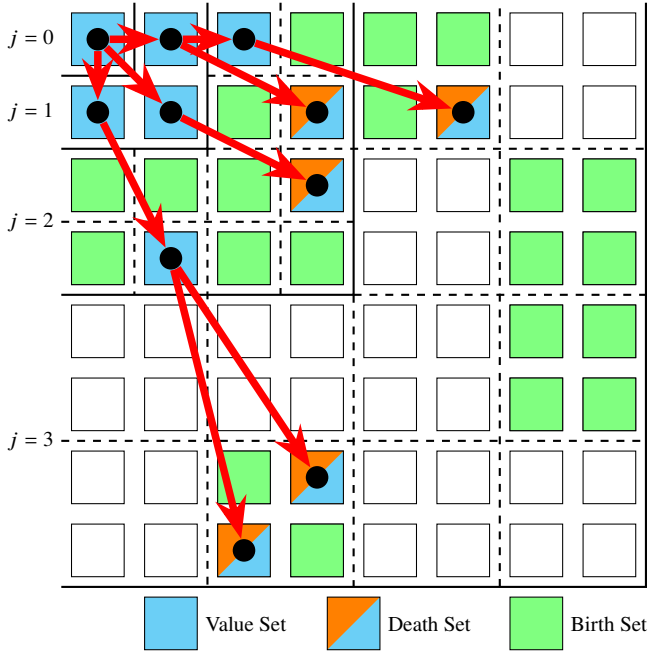
\begin{figure}
    \centering
    \begin{scaletikzpicturetowidth}{\columnwidth}
        \begin{tikzpicture}[scale=\tikzscale,
                            squarednode/.style={rectangle,draw,minimum size=7mm},
                            treebranch/.style={-Stealth,color=red,line width=1mm},
                            treenode/.style={circle,minimum size=3mm,inner sep=0pt,fill=black},
                            fill lower right/.style={path picture={\fill[#1] (path picture bounding box.south west) -- (path picture bounding box.north east) |-cycle;}}]

            \def\value{cyan!50}
            \def\death{orange}
            \def\birth{green!50}

            \node [squarednode,fill=\value,label=right:{Value Set}] at (1.5,-0.5) {};
            \node [squarednode,fill=\death,fill lower right=\value,label=right:{Death Set}] at (4.0,-0.5) {};
            \node [squarednode,fill=\birth,label=right:{Birth Set}] at (6.5,-0.5) {};

            \node [squarednode,fill=\value] at (0.5,7.5) {};

            \node [squarednode,fill=\value] at (0.5,6.5) {};
            \node [squarednode,fill=\value] at (1.5,6.5) {};
            \node [squarednode,fill=\value] at (1.5,7.5) {};

            \foreach \y in {4,5}{
                \foreach \x in {0,1}
                    \node [squarednode] at (\x+0.5,\y+0.5) {};
                \foreach \x in {2,3}
                    \node [squarednode] at (\x+0.5,\y+0.5) {};
                }
            \foreach \x in {2,3}
                \foreach \y in {6,7}
                    \node [squarednode] at (\x+0.5,\y+0.5) {};
            \node [squarednode,fill=\value] at (1.5,4.5) {};
            \node [squarednode,fill=\value] at (2.5,7.5) {};
            \node [squarednode,fill=\death,fill lower right=\value] at (3.5,5.5) {};
            \node [squarednode,fill=\death,fill lower right=\value] at (3.5,6.5) {};
            \node [squarednode,fill=\birth] at (0.5,4.5) {};
            \node [squarednode,fill=\birth] at (0.5,5.5) {};
            \node [squarednode,fill=\birth] at (1.5,5.5) {};
            \node [squarednode,fill=\birth] at (2.5,5.5) {};
            \node [squarednode,fill=\birth] at (2.5,6.5) {};
            \node [squarednode,fill=\birth] at (2.5,4.5) {};
            \node [squarednode,fill=\birth] at (3.5,4.5) {};
            \node [squarednode,fill=\birth] at (3.5,7.5) {};

            \foreach \y in {0,...,3}{
                \foreach \x in {0,...,3}
                    \node [squarednode] at (\x+0.5,\y+.5) {};
                \foreach \x in {4,...,7}
                    \node [squarednode] at (\x+0.5,\y+.5) {};
            }
            \foreach \x in {4,...,7}
                \foreach \y in {4,...,7}
                    \node [squarednode] at (\x+0.5,\y+.5) {};
            \node [squarednode,fill=\death,fill lower right=\value] at (2.5,0.5) {};
            \node [squarednode,fill=\death,fill lower right=\value] at (3.5,1.5) {};
            \node [squarednode,fill=\death,fill lower right=\value] at (5.5,6.5) {};
            \node [squarednode,fill=\birth] at (4.5,7.5) {};
            \node [squarednode,fill=\birth] at (5.5,7.5) {};
            \node [squarednode,fill=\birth] at (4.5,6.5) {};
            \node [squarednode,fill=\birth] at (2.5,1.5) {};
            \node [squarednode,fill=\birth] at (3.5,0.5) {};
            \foreach \x in {6,7}
                \foreach \y in {2,...,5}
                    \node [squarednode,fill=\birth] at (\x+0.5,\y+.5) {};

            \draw[thick] (1,8) -- (1,7) -- (0,7);
            \draw[thick] (2,8) -- (2,6) -- (0,6);
            \draw[thick] (4,8) -- (4,4) -- (0,4);
            \draw[thick] (8,8) -- (8,0) -- (0,0);

            \draw[dashed,thick] (1,6) -- (1,4);
            \draw[dashed,thick] (2,6) -- (2,0);
            \draw[dashed,thick] (3,8) -- (3,4);
            \draw[dashed,thick] (4,4) -- (4,0);
            \draw[dashed,thick] (0,2) -- (8,2);
            \draw[dashed,thick] (6,0) -- (6,8);
            \draw[dashed,thick] (2,6) -- (8,6);
            \draw[dashed,thick] (4,4) -- (8,4);
            \draw[dashed,thick] (0,5) -- (4,5);
            \draw[dashed,thick] (2,7) -- (4,7);

            \coordinate[treenode] (t0) at (0.5,7.5);
            \coordinate[treenode] (t1) at (0.5,6.5);
            \coordinate[treenode] (t2) at (1.5,6.5);
            \coordinate[treenode] (t3) at (1.5,7.5);
            \coordinate[treenode] (t4) at (2.5,7.5);
            \coordinate[treenode] (t5) at (3.5,6.5);
            \coordinate[treenode] (t6) at (5.5,6.5);
            \coordinate[treenode] (t7) at (3.5,5.5);
            \coordinate[treenode] (t8) at (1.5,4.5);
            \coordinate[treenode] (t9) at (3.5,1.5);
            \coordinate[treenode] (t10) at (2.5,0.5);

            \draw[treebranch] (t0) -- (t1);
            \draw[treebranch] (t0) -- (t2);
            \draw[treebranch] (t0) -- (t3);
            \draw[treebranch] (t3) -- (t4);
            \draw[treebranch] (t3) -- (t5);
            \draw[treebranch] (t4) -- (t6);
            \draw[treebranch] (t2) -- (t7);
            \draw[treebranch] (t1) -- (t8);
            \draw[treebranch] (t8) -- (t9);
            \draw[treebranch] (t8) -- (t10);

            \node[left] () at (0,7.5) {\(j=0\)};
            \node[left] () at (0,6.5) {\(j=1\)};
            \node[left] () at (0,5) {\(j=2\)};
            \node[left] () at (0,2) {\(j=3\)};

        \end{tikzpicture}
    \end{scaletikzpicturetowidth}
    \caption[Trans-dimensional proposal sets]{Proposal sets for the 2D wavelet tree parameterisation.
    The wavelet coefficients are arranged as in \autoref{fig:wavtree}, although here they are coloured by the proposal that may be performed at that coefficient.
    Value, death and birth proposals can be made at the blue, orange and green coefficients, respectively.
    Note that the death set is a subset of the value set, and the birth set is disjoint from the other two sets.}
    \label{fig:sets}
\end{figure}

In this birth-death sampler we have three classes of proposals: (1) value proposals, where the value of a wavelet coefficient is modified; (2) birth proposals, where a new coefficient is added to the tree; and (3) death proposals, where a coefficient is removed from the tree.
We divide up our wavelet coefficient space into three sets, \(\mathcal{S}_v, \mathcal{S}_b, \mathcal{S}_d\), which contain the coefficients at which a value, birth or death proposal, respectively, can be performed.
An example is shown in \autoref{fig:sets}.
The value set \(\mathcal{S}_v\) corresponds to the current tree, and will always include at least the root of the tree, hence \(|\mathcal{S}_v| \ge 1\).
The death set \(\mathcal{S}_d\) is the ``low-hanging fruit'' of the tree, i.e.\ coefficients without any active child coefficients.
The birth set \(\mathcal{S}_b\) is all the currently inactive child coefficients of the current tree.
Notice that the death set is a proper subset of the value set, \(\mathcal{S}_d \subset \mathcal{S}_v\), and the birth set is disjoint from the other two sets, \(\mathcal{S}_b \cap \mathcal{S}_v = \mathcal{S}_b \cap \mathcal{S}_d = \varnothing\).

The value proposal distribution is given by
\begin{equation}
    q_v(\kappa'|\kappa) = \frac{q(\Delta\varkappa_i|i)}{|\mathcal{S}_v|},
    \label{eqn:prop_value}
\end{equation}
where the numerator is the distribution from which a perturbation to wavelet coefficient \(\varkappa_i\) is drawn.
We choose this to be a zero-mean Normal distribution with variance tuned to achieve an acceptance rate of roughly 20--40\%.
The denominator covers the probability of choosing to perturb the \(i\)\textsuperscript{th} coefficient.

Birth proposals are performed by selecting a coefficient \(i\) from the birth set \(\mathcal{S}_b\) and initialising it with a value drawn from the value prior.
In the case where the birth set is empty, i.e.\ the current tree spans all coefficients, the proposal probability is zero.
Hence, we have
\begin{equation}
    q_b(\kappa'|\kappa) = \frac{p(\varkappa_i|\mathcal{T}_k,k)}{|\mathcal{S}_b|}.
    \label{eqn:prop_birth}
\end{equation}

Death proposals simply involve choosing a coefficient \(i\) to remove from the death set \(\mathcal{S}_d\), so
\begin{equation}
    q_d(\kappa'|\kappa) = \frac{1}{|\mathcal{S}_d|}.
    \label{eqn:prop_death}
\end{equation}

The acceptance criterion (\autoref{eqn:alpha}) requires the reverse proposal distributions for each of these proposals \(q(\kappa|\kappa')\).
The value proposal distribution is symmetric, so the reverse value proposal is equal to the forward proposal.
As for the birth and death proposals, these are the reverse proposals of each other, i.e.\ the reverse of proposing the birth of coefficient \(i\) from \(\mathcal{S}_b\) is proposing its death from \(\mathcal{S}_d'\), where the prime here denotes the set after the proposal.
So the reverse proposals are
\begin{align}
    q_v(\kappa|\kappa') &= q_v(\kappa'|\kappa), \\
    q_b(\kappa|\kappa') &= q_d(\kappa'|\kappa), \\
    q_d(\kappa|\kappa') &= q_b(\kappa'|\kappa). 
\end{align}

\subsection{Acceptance criteria}
Recall the acceptance criterion (\autoref{eqn:alpha}) depends on the prior, likelihood and proposal ratios, and a Jacobian.
The likelihood ratio remains the same regardless of the proposal, and we denote this as \(\mathcal{L}(\kappa', \kappa) = p(\gamma'|\kappa')\;/\;p(\gamma|\kappa)\).
The prior and proposal ratios depend on the type of proposal performed.

For value proposals the tree structure does not change, so the prior ratio is simply whether or not the proposed coefficient value is more likely than the current value
\begin{equation}
    \frac{p(\kappa')}{p(\kappa)} = \frac{p(\varkappa_i'|\mathcal{T}_k,k)}{p(\varkappa_i|\mathcal{T}_k,k)}.
\end{equation}
As for the proposal ratio, the reverse distribution is equal to the forward distribution so the ratio is 1.
Since there is no change in dimension of the tree, the Jacobian will always be the identity, with determinant equal to 1.
Hence the acceptance criterion for value proposals is
\begin{equation}
    \alpha_v(\kappa',\kappa) = \min\left\{1, \frac{p(\varkappa_i'|\mathcal{T}_k,k)}{p(\varkappa_i|\mathcal{T}_k,k)}\mathcal{L}(\kappa', \kappa)\right\}.
    \label{eqn:value_acceptance}
\end{equation}

For birth proposals, the tree structure changes.
Using our uniform prior for the number of coefficients in the tree, this part of the prior ratio is unity, \(p(k+1)\;/\;p(k) = 1\).
The prior values also all cancel out except at the proposed new coefficient \(i\).
So the overall birth prior ratio is, dropping the conditional dependencies for clarity
\begin{equation}
    \frac{p(\kappa')}{p(\kappa)} = \frac{p(\mathcal{T}_{k+1})p(\varkappa_i)}{p(\mathcal{T}_k)}.
\end{equation}
Using reverse distributions above, the proposal ratio is
\begin{equation}
    \frac{q(\kappa|\kappa')}{q(\kappa'|\kappa)} = \frac{|\mathcal{S}_b|}{|\mathcal{S}_d'|p(\varkappa_i)}.
\end{equation}
For the Jacobian, it is a simple case of \citet{Green1995}'s dimension matching without any transformation of random variables.
We can denote our parameter space \(\theta\) as a set of \(k\) tuples each containing a unique index \(t_i\) denoting a position in the tree and the value of the wavelet coefficient at that position \(\varkappa_i\).
Dimension matching then requires us to sample random variables \(u\) and \(w\) for the position and value of the new coefficient, respectively, and then setting the proposal \(\theta'\) to be some function of \(u\) and \(w\).
Using the proposals defined previously, that function is simply the identity.
For all the pre-existing parameter (\(i=1\dots k\)) there is no change, and for the new coefficient we set \((t_{k+1}, \varkappa_{k+1}) = (u, v)\), hence there is no transformation of the random variable and the Jacobian is the identity matrix.
So the birth proposal acceptance criterion is
\begin{equation}
    \alpha_b(\kappa',\kappa) = \min\left\{1, \frac{p(\mathcal{T}_{k+1})}{p(\mathcal{T}_k)}\frac{|\mathcal{S}_b|}{|\mathcal{S}_d'|}\mathcal{L}(\kappa', \kappa)\right\}.
    \label{eqn:birth_acceptance}
\end{equation}

Similar reasoning and considering that the Jacobian of the death proposal is the inverse of the birth proposal Jacobian leads to the death proposal acceptance criterion, 
\begin{equation}
    \alpha_d(\kappa',\kappa) = \min\left\{1, \frac{p(\mathcal{T}_{k-1})}{p(\mathcal{T}_k)}\frac{|\mathcal{S}_d|}{|\mathcal{S}_b'|}\mathcal{L}(\kappa', \kappa)\right\}.
    \label{eqn:death_acceptance}
\end{equation}

\subsection{The full trans-dimensional MCMC algorithm}
Starting from a randomly initialised convergence model \(\kappa_0\) with \(k=1\), trans-dimensional MCMC iterates for N iterations, for sufficiently large N.
At each iteration \(t\), the type of proposal is chosen with probability \(p(\mathrm{birth})\), \(p(\mathrm{death})\) and \(p(\mathrm{value})\).
These probabilities can be set arbitrarily, subject to the constraints that \(p(\mathrm{birth}) = p(\mathrm{death})\) and \(p(\mathrm{birth}) + p(\mathrm{death}) + p(\mathrm{value}) = 1\) \citep{Hawkins2015}.
Having selected the proposal type, a new model \(\kappa'\) is proposed according to corresponding proposal distribution \(q(\kappa'|\kappa_t)\) (\hyperref[eqn:prop_value]{Equations~\ref{eqn:prop_value},~\ref{eqn:prop_birth}~and~\ref{eqn:prop_death}}).
The proposal is then accepted with probability \(\alpha(\kappa', \kappa_t)\) according to the corresponding acceptance criterion (\hyperref[eqn:value_acceptance]{Equations~\ref{eqn:value_acceptance},~\ref{eqn:birth_acceptance}~and~\ref{eqn:death_acceptance}}) by setting the next sample in the chain \(\kappa_{t+1}=\kappa'\).
Note that at each iteration, only a single model parameter (wavelet coefficient \(\varkappa_i\)) is affected.
In practice, the first \(N_{\mathrm{burn}}\) iterations are discarded and only every \(N_{\mathrm{thin}}\)\textsuperscript{th} sample thereafter is kept.
This avoids saving too many initial samples that are far from the posterior peak and reducing the correlation between samples, while still allowing for adequate sampling of the posterior.
\hyperref[algo:TDT]{Algorithm~\ref{algo:TDT}} summarises the trans-dimensional MCMC algorithm.
\begin{algorithm}
    \DontPrintSemicolon{}
    \caption{Trans-dimensional (Birth/Death) MCMC algorithm}\label{algo:TDT}
    \KwIn{Number of iterations \(N\)}
    \KwData{Data vector \(\gamma\in\mathbb{C}^M\)}
    \KwResult{Chain of length \(N\) of model parameter vectors \(\kappa\in\mathbb{R}^M\)}
    \Begin{
        \(t \leftarrow 0\)\;
        Randomly choose initial sample \(\kappa_0\)\;
        \While{\(t < N\)}{
            Draw random number \(u \sim U(0,1)\)\;
            \uIf{\(u < p(\mathrm{birth})\)}{
                Proposal type is birth\;
            }
            \uElseIf{\(u < 2p(\mathrm{birth})\)}{
                Proposal type is death\;
            }
            \Else{
                Proposal type is value\;
            }
            Propose new sample \(\kappa'\) from corresponding \(q(\kappa'|\kappa_t)\)\;
            Calculate acceptance probability corresponding \(\alpha(\kappa',\kappa_t)\)\;
            Draw random number \(u \sim U(0,1)\)\;
            \If{\(u > \alpha\)}{
                Reject proposal \(\kappa'\)\;
            }
            \Else{
                Accept proposal \(\kappa'\)\;
                \(\kappa_{t+1} \leftarrow \kappa'\)\;
            }
            \(t \leftarrow t+1\)\;
        }
    }
\end{algorithm}

\section{Mass-mapping with Trans-dimensional Trees}
\label{sec:tdtmm}
In this section we give details of how we use the trans-dimensional tree structure described previously for mass-mapping.
In particular we describe the forward operator and choice of prior used for the following sections.

\subsection{Forward Operator}
Our trans-dimensional MCMC samples the wavelet coefficients \(\varkappa\) of a convergence map \(\kappa\).
The convergence map is then \(\kappa = W^{-1}\varkappa\), where \(W^{-1}\) is the inverse wavelet transform.
To use the KS kernel in \autoref{eqn:KS_tdt}, denoted by \(D\) we need forward and inverse fast Fourier transforms, \(F, F^{-1}\), respectively.
As such, obtaining the shear from sampled convergence wavelet coefficients is given by
\begin{equation}
    \label{eqn:fwd}
    \gamma = F^{-1}DFW^{-1}\varkappa.
\end{equation}
The input data for our inversions are gridded galaxy shear measurements.
Gridding is necessary to allow the galaxy ellipticity due to shear to emerge out of the shape noise due to intrinsic ellipticity.
We make the common assumption that the intrinsic ellipticities of galaxies are Gaussian randomly oriented with zero mean, and thus we take the observed shear in a small region of sky (pixel) to be the mean of the ellipticities of the galaxies in that region.
However, as a result of the gridding we are left with fewer data points, and likely insufficient data to produce a high-resolution convergence map.
Further, depending on the coarseness of the gridding it is possible that some pixels contain no galaxy measurements, as is common in weak lensing surveys, which may lead to signal leakage between Fourier modes.

We assume that the shear measurements in each pixel follow a Gaussian distribution of known variance \(\sigma_i^2\).
As such we set the likelihood of observing shear measurements \(\gamma\) for a given convergence map wavelet coefficients \(\varkappa\) to be
\begin{equation}
    p(\gamma|\varkappa) = \frac{1}{\sqrt{(2\pi)^N|C_d|}}\exp{
        \left\{
            \frac{
                (\Phi\varkappa - \gamma)^T C_d^{-1} (\Phi\varkappa - \gamma)
                }
            {2}
        \right\}
        },
\end{equation}
where \(\Phi\) is the operator defined in \autoref{eqn:fwd} and \(C_d\) is the data covariance matrix.
We typically assume no correlation between neighbouring pixels, and as such \(C_d\) is a diagonal matrix with all the \(\sigma_i^2\) along the diagonal.
We also assume that the real and imaginary parts of the \(\sigma_i^2\) are equal.
We note that these assumptions can break down in the presence of systematics such as intrinsic alignment.

\subsection{Generalised Gaussian Prior on Wavelet Coefficients}
\label{subsec:ggd}
\begin{figure}
    \includegraphics[width=\columnwidth]{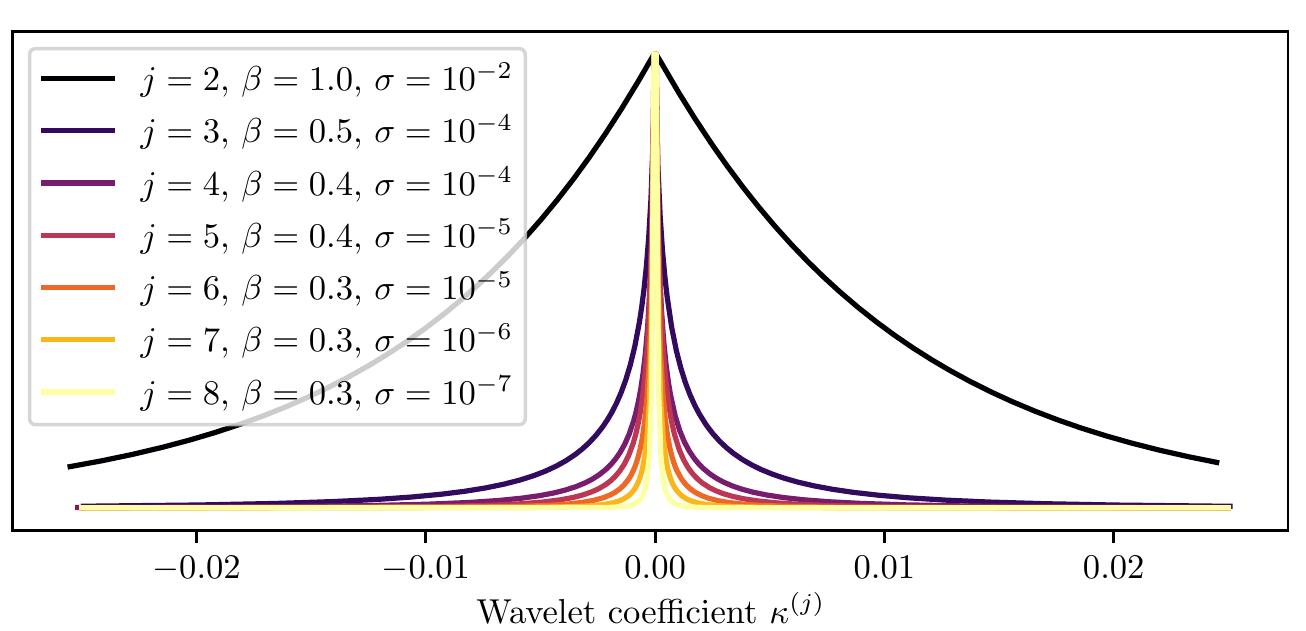}
    \caption[Generalised Gaussian distribution of wavelet coefficients of mass-maps]{Fitted generalised Gaussian distributions of wavelet coefficients for the Bolshoi 7 \& 8 N-body galaxy cluster simulations, normalised to have unit height.
    The scaling function coefficients and the largest wavelet scale coefficients \(j=1\) have been omitted, as there are not enough coefficients for a fit to be truly meaningful.
    The fitted scale, \(\sigma\), and shape, \(\beta\) parameters are indicated.}
    \label{fig:wavs_ggds}
\end{figure}

Strictly, sparsity is measured by the \(\ell_0\)-norm, which counts the non-zero elements in a vector.
However, this norm is non-convex, and it has been shown that the convex \(\ell_1\)-norm, or equivalently, the Laplace distribution, produces comparable results, and so is now widely used for sparsity promotion \citep{Donoho2006,Candes2011}.
However, thanks to the multiresolution nature of the wavelets we can impose different priors for different length scales.
In particular, we expect the wavelet coefficients of larger scale structure to be more Gaussian than at smaller scales.
\citet{Starck2021} had a similar idea and solved for sparsity-based component to capture non-Gaussian structure and a separate Gaussian random field.
Alternatively, \citet{McEwen2017} suggested fitting a generalised-Gaussian distribution (GGD) to each wavelet scale.
The probability density function of a GGD is given by
\begin{equation}
    \label{eqn:ggd}
    p(x|\mu,\sigma,\beta) = \frac{\beta}{2\sigma\Gamma(\beta^{-1})}\exp\left\{-\left|\frac{x-\mu}{\sigma}\right|^{\;\beta}\right\},
\end{equation}
where \(\Gamma(\cdot)\) is the gamma function, \(\mu\) is the mean, \(\sigma\) is the scale parameter (c.f.\ standard deviation) and \(\beta\) is the shape parameter.
For \(\beta = 2\) this gives the Gaussian distribution, \(\beta = 1\) gives the Laplacian, and as \(\beta\rightarrow \infty\) this tends to a uniform distribution in the range \([\mu-\sigma, \mu+\sigma]\).
\citet{McEwen2017} suggested learning the scale and shape parameters for each wavelet scale based on simulation images.
\autoref{fig:wavs_ggds} shows the fitted GGDs of wavelet coefficients at each scale for the Bolshoi 7 \& 8 galaxy cluster simulation convergence images \citep{Klypin2011}.
The fitted scale and shape parameters are also shown, and clearly the larger scale wavelet coefficients are more Gaussian than the smaller scale coefficients.
We use the distributions in \autoref{fig:wavs_ggds} as loose guide for tuning the shape and scale parameters for each wavelet scale for the application to simulation data in the following section.
It would be unreasonable to use the exact distributions as they would not be known in the real data case.
We note that this prior implies that the individual wavelet coefficients are considered independent.

In the wavelet tree parameterisation, the root of the tree is a single pixel approximation of the \(\kappa\) map, i.e.\ the image mean.
However, the mass-sheet degeneracy means \(\kappa\) can only be determined up to an unknown constant \(\lambda\), \(\kappa_{\lambda} = \lambda\kappa + (1 - \lambda)\).
We can then set the prior of the tree root to be extremely tight around 0, as it cannot be determined from the data.

\section{Application to simulations}
\label{sec:sims}
In this section we present the results of tests of our trans-dimensional tree method on datasets generated from simulated cluster-scale convergence maps. 

\subsection{Data}
We begin with a ground truth convergence map \(\kappa\), extracted from the Bolshoi N-body simulation \citep{Klypin2011}.
We use the same simulation catalogues as \citet{Lanusse2016} and \citet{Price2021}, extracted using the CosmoSim\footnote{\url{http://www.cosmosim.org}} website.
This assumed a redshift of \(z=0.3\), a \(10\times10\) arcmin\textsuperscript{2} field of view, and the convergence maps have been normalised with respect to lensing sources at infinity.
The ground truth image is shown in \autoref{fig:bol8}, sampled on a \(256\times256\) pixel grid.

From the ground truth convergence maps we generate noisy synthetic shear data \(\gamma\) from
\begin{equation}
    \gamma = F^{-1}DF\kappa + n,
\end{equation}
where \(n\) is drawn from a zero-mean Gaussian distribution \(\mathcal{N}(0, \sigma^2)\), \(\sigma^2\) is the data covariance determined by the number of observed galaxies in a pixel and an assumed intrinsic galaxy ellipticity dispersion of 0.37 \citep{Price2021},
\begin{equation}
    \sigma^2 = \frac{0.37^2}{\sqrt{2N}}
\end{equation}
where \(N\) is the expected number of galaxies in a pixel for a given number density of galaxies per arcmin\textsuperscript{2}, \(n_{\mathrm{gal}}\).
For example, \emph{Euclid} is expected to be able to see about \(n_{\mathrm{gal}}\sim30\) galaxies per arcmin\textsuperscript{2} \citep[e.g][]{Laureijs2011}.
To make the data somewhat more realistic, we apply a random mask to the noisy shear data whereby we mask out 1\% of pixels to simulate regions with no galaxy observations.

\subsection{Method}
We invert the synthetic shear data for a convergence map using the trans-dimensional tree method detailed in \autoref{sec:tdt}.
The wavelets we use are the Cohen-Debauchies-Feauveau 9/7 wavelet family \citep{Cohen1992}, as these produced the best results for \citet{Hawkins2015}.
We run a single Markov chain for at least \(10^6\) steps and check that the likelihood and number of model parameters \(k\) has converged.
If needed we restart the chain from where it ended and extend the chain for as long as necessary.
The generalised Gaussian prior parameters (shape and scale) are tuned for each wavelet scale, except the lowest scale which represents the mass-sheet degeneracy, so as to achieve an appropriate acceptance rate, typically around 20--50\%.
Thus these are not free parameters during inference.
Again, the values obtained by fitting a GGD to the ground truth Bolshoi simulations (\autoref{fig:wavs_ggds}) are used as a loose guide, and we ensure that our priors are wider than these fitted distributions.
It is conceivable that these parameters be determined in a hierarchical manner \citep[e.g.][]{Bodin2012,Alsing2016}, however we leave this for future work.
As in \citet{Price2021} we evaluate the similarity between our convergence solutions (mean or best-fitting) and the ground truth using two quantities: the signal-to-noise ratio (SNR) in dB to assess the overall difference
\begin{equation}
    \mathrm{SNR}(\kappa^*) = 10 \times \log_{10}\left(\frac{\|\kappa\|_2^2}{\|\kappa - \kappa^*\|_2^2}\right)
\end{equation}
and the Pearson correlation coefficient (\(r\)) to assess the structural similarities.
\begin{equation}
    r(\kappa^*) = \frac{\sum_{i=1}^M(\kappa_i - \langle\kappa\rangle)(\kappa^*_i - \langle\kappa^*\rangle)}{\sqrt{\sum_{i=1}^M(\kappa_i - \langle\kappa\rangle)^2}\sqrt{\sum_{i=1}^M(\kappa^*_i - \langle\kappa^*\rangle)^2}}
\end{equation}
Here \(\kappa^*\) denotes our solution convergence map and \(\langle\cdot\rangle\) denotes an average over the \(M\) pixels.

\subsection{Results}
\subsubsection{Recovery of a clean high-resolution mass-map}
\begin{figure}
    \centering
    \includegraphics[width=\columnwidth]{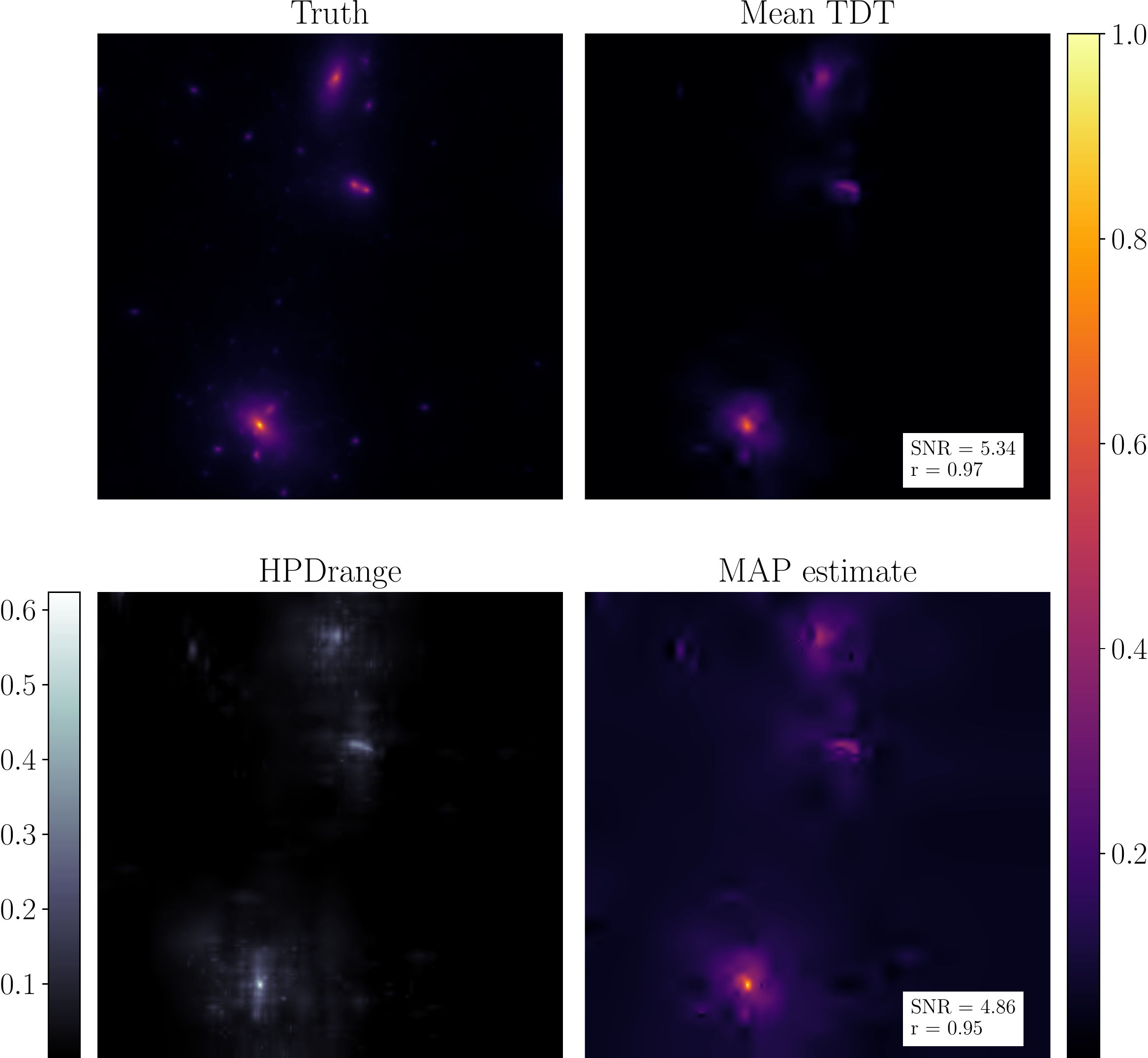}
    \caption[Bolshoi simulation results]{(top left) Ground truth convergence field from which synthetic shear measurements were obtained.
    (top right) Mean solution from trans-dimensional MCMC\@.
    (bottom right) Highest posterior sample (estimate of the MAP solution) from trans-dimensional MCMC\@.
    (bottom left) Size of the highest posterior density region as a measure of uncertainty.
    All images show a \(10\times10\) arcmin\textsuperscript{2} field of view.}
    \label{fig:bol8}
\end{figure}

\autoref{fig:bol8} shows the ground truth map, the mean and highest posterior (MAP estimate) trans-dimensional MCMC point solutions and the range of the highest posterior density region at the 99\% credible interval level.
This is a very clean data example.
Both the mean and highest posterior point estimates recover the three main high convergence regions well, although some of the fainter sources are missing, possibly lost in the noise or smoothed out when averaging over many samples in the case of the mean solution.
The highest posterior point estimate recovers nicely the peak at the core of the lowermost cluster, which seems to have been lost in the mean solution.
In terms of the overall SNR and correlation, however, the mean solution is better than the MAP\@.

At a given pixel, the quantified uncertainty is represented by the two-tailed 99\% credible interval range, symmetric about the peak of the histogram of values that the pixel takes in our MCMC chain.
Assuming a unimodal histogram, this is the highest posterior density region (HPDRange in \autoref{fig:bol8}).
The Bayesian credible interval is an interval in which the pixel value falls with probability 99\%.
The credible interval range is thus calculated as the range between the 0.5\textsuperscript{th} and 99.5\textsuperscript{th} percentiles of individual pixel values of the convergence map.
Credible intervals are widely used Bayesian measures of uncertainty \citep[e.g.][]{Gelman2013,Pereyra2017,Price2020,Marignier2023}.
The lateral extent of the three main clusters in the uncertainty map are slightly larger than those in the point solutions.
As well as highlighting the three main peaks, where the detailed structures are not quite resolved, the uncertainty map picks out some of the smaller, fainter structures that have been missed in the point solutions.
For example, a faint source in the top left quadrant is identified, as are faint structures in the vicinity of topmost and lowermost main peaks.
We emphasise here that while we have chosen to show a credible interval as our measure of uncertainty for its Bayesian interpretation, with adequate sampling of the posterior any measure of uncertainty can be obtained.

\begin{figure*}
    \includegraphics[width=\textwidth]{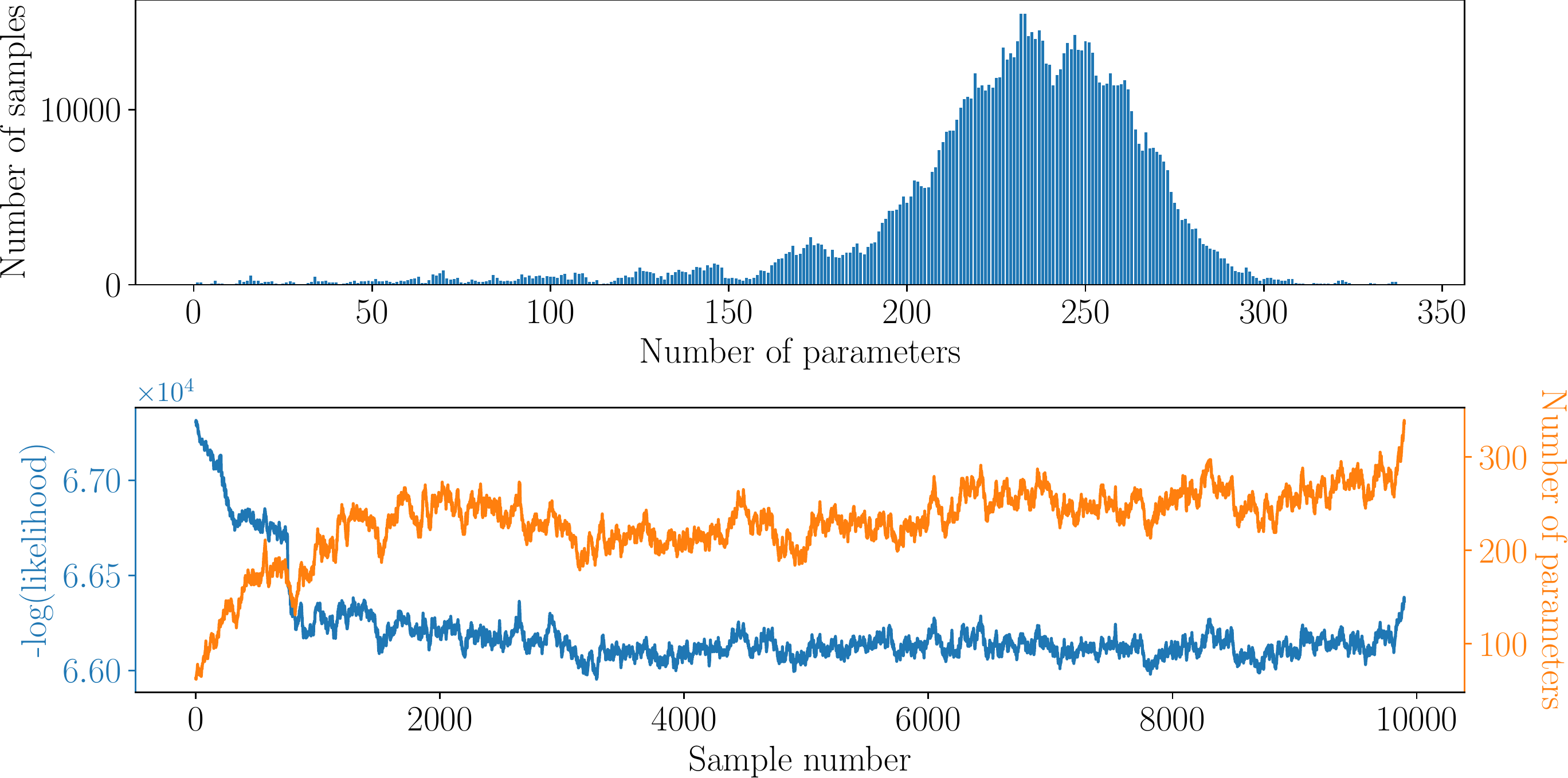}
    \caption[Inversion history]{History of the MCMC chain that produces the results in \autoref{fig:bol8}.
    (top) Histogram of number of wavelet coefficients.
    (bottom) Evolution over the Markov chain of the likelihood (blue) and number of wavelet coefficients (orange).
    Note that the samples in the bottom plot have been thinned, i.e.\ only every 100\textsuperscript{th} sample is shown, but have not been thinned in the top plot, i.e.\ every sample is shown.
    This explains why the top plot contains more samples (area under the histogram) than the bottom plot (length of \(x\) axis).}
    \label{fig:tdt_history}
\end{figure*}

\autoref{fig:tdt_history} shows a histogram of the number of active tree nodes (non-zero wavelet coefficients, \(k\)) for all the samples in the trans-dimensional MCMC chain, as well as the evolution of the likelihood \(p(\gamma|\kappa)\) and the number of wavelet coefficients \(k\) as the chain progresses.
The number of coefficients generally increases, as designed by the parameterisation.
As the number of coefficients increases, the likelihood expectedly improves.
However, the number of parameters does not increase indefinitely and converges at around 250 coefficients.
In this case, the ground truth is sampled on a \(256\times256\) pixel grid, meaning the maximum potential size of the parameter space is \(k_{\max} =\;\)\num{65536}.
This is a massive reduction of the parameter space.

\subsubsection{Low-resolution and high-noise inversions}
In the example shown previously, the noise level was kept relatively low, using \(n_{gal} =\;\)\num{5000} galaxies per arcmin\textsuperscript{2} which corresponds to just under 8 galaxies per pixel.
The \emph{Euclid} telescope is expecting to observe around 30 galaxies per arcmin\textsuperscript{2} of sky \citep{Laureijs2011}.
For this \(10\times10\) arcmin patch of sky, that would correspond to on average less than 0.05 galaxies per pixel.
During our initial experiments, we found this noise level to be too high for our method to work.
Adding or removing parameters would cause so little a difference to the likelihood (data fit) that almost any proposal would be accepted, and since births and deaths are proposed in the same proportions, the tree would never grow.

\begin{figure*}
    \includegraphics[width=\textwidth]{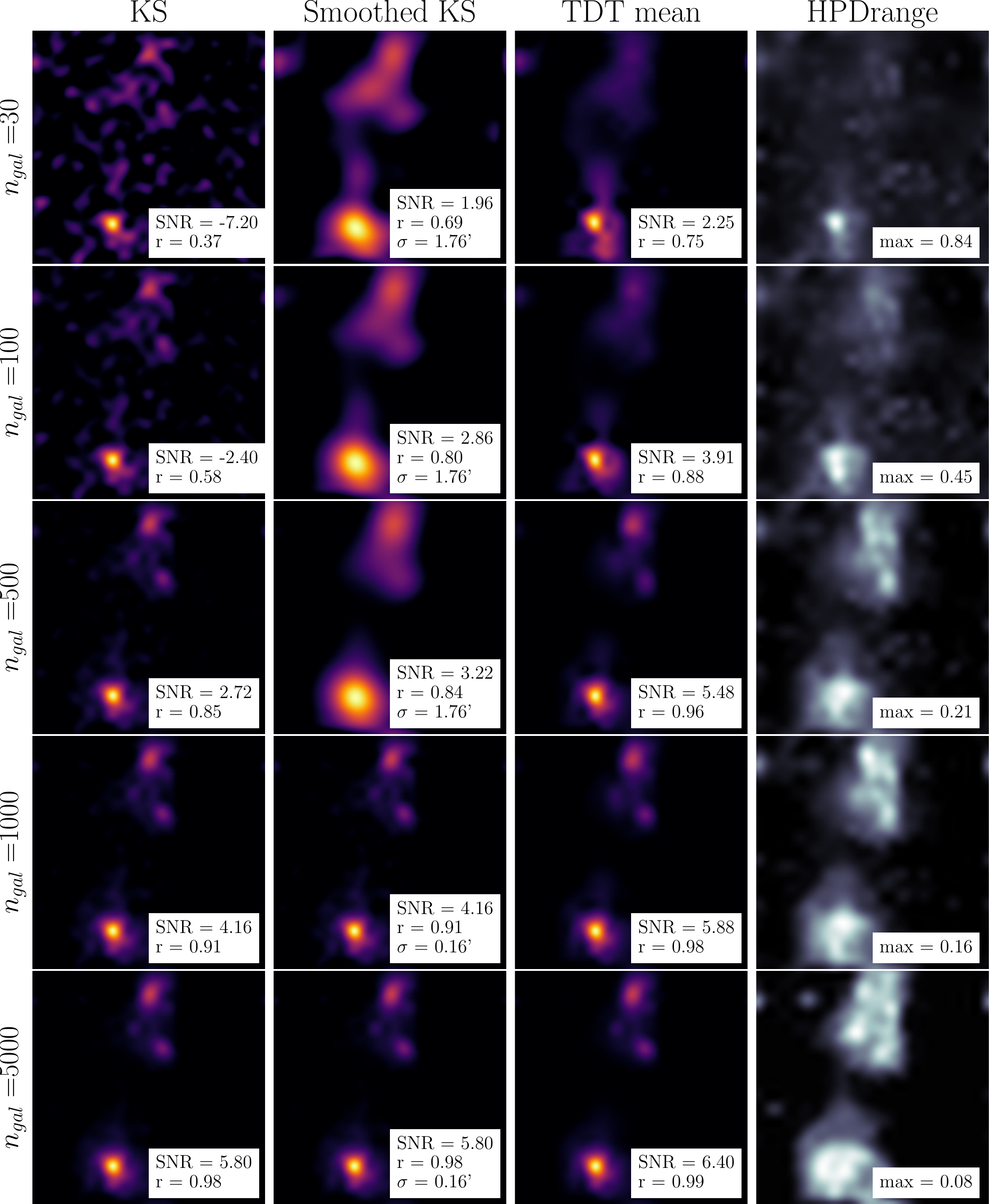}
    \caption[Comparison of trans-dimensional trees against KS]{Comparison of trans-dimensional trees against KS at various levels of noise.
    All images show a \(10\times10\) arcmin\textsuperscript{2} field of view.
    The ground truth image is the same as that in \autoref{fig:bol8}, resampled onto a \(32\times32\) grid to beat down the noise by increasing the number of average galaxies per pixel.
    Each row is for a given noise level, defined by an average number of galaxies per arcmin\textsuperscript{2}, \(n_{gal}\).
    Noise decreases from top to bottom.
    Each column shows, from left to right, the KS reconstruction, the KS reconstruction with optimum smoothing, the mean of the trans-dimensional MCMC, and the uncertainty produced from MCMC\@.
    The quality of the reconstructions in the first three columns is indicated by the SNR and correlation coefficient.
    The optimum Gaussian smoothing kernel size \(\sigma\) in arcmin is also indicated (second column), as is maximum uncertainty value (last column).}
    \label{fig:ks_comparison}
\end{figure*}

The solution that we found to beat down the noise is to decrease the resolution.
At \(32\times32\) pixels, a \emph{Euclid}-like noise level would correspond to about 3 galaxies per pixel for this patch of sky.
\autoref{fig:ks_comparison} shows the results (mean and uncertainty) of our trans-dimensional inversions at this lower resolution for varying noise levels.
Also shown are the KS solution and the KS solution with optimum smoothing.
The smoothing uses a simple 2D Gaussian kernel, the optimum size of which was determined by a simple linear search of the kernel size \(\sigma\) and identifying which kernel produced the best SNR\@.
This would obviously not be possible in practice with real data, as it requires knowledge of the true convergence field.
As such, this optimum solution represents the top-end of what could be possible with KS\@.
At all noise levels, trans-dimensional MCMC performs better than the optimally smoothed KS solution in terms of SNR and correlation with the ground truth.
Visually, even at the highest noise levels, the lowermost peak is more tightly constrained by MCMC\@.
The additional uncertainty information obtained from MCMC provides further constraints on the locations of peaks, although, somewhat surprisingly, the uncertainty is more laterally spread in the low noise cases than in the high noise cases.
The size of the uncertainty does decrease as the noise as noise decreases, as expected.

\begin{figure*}
    \includegraphics[width=\textwidth]{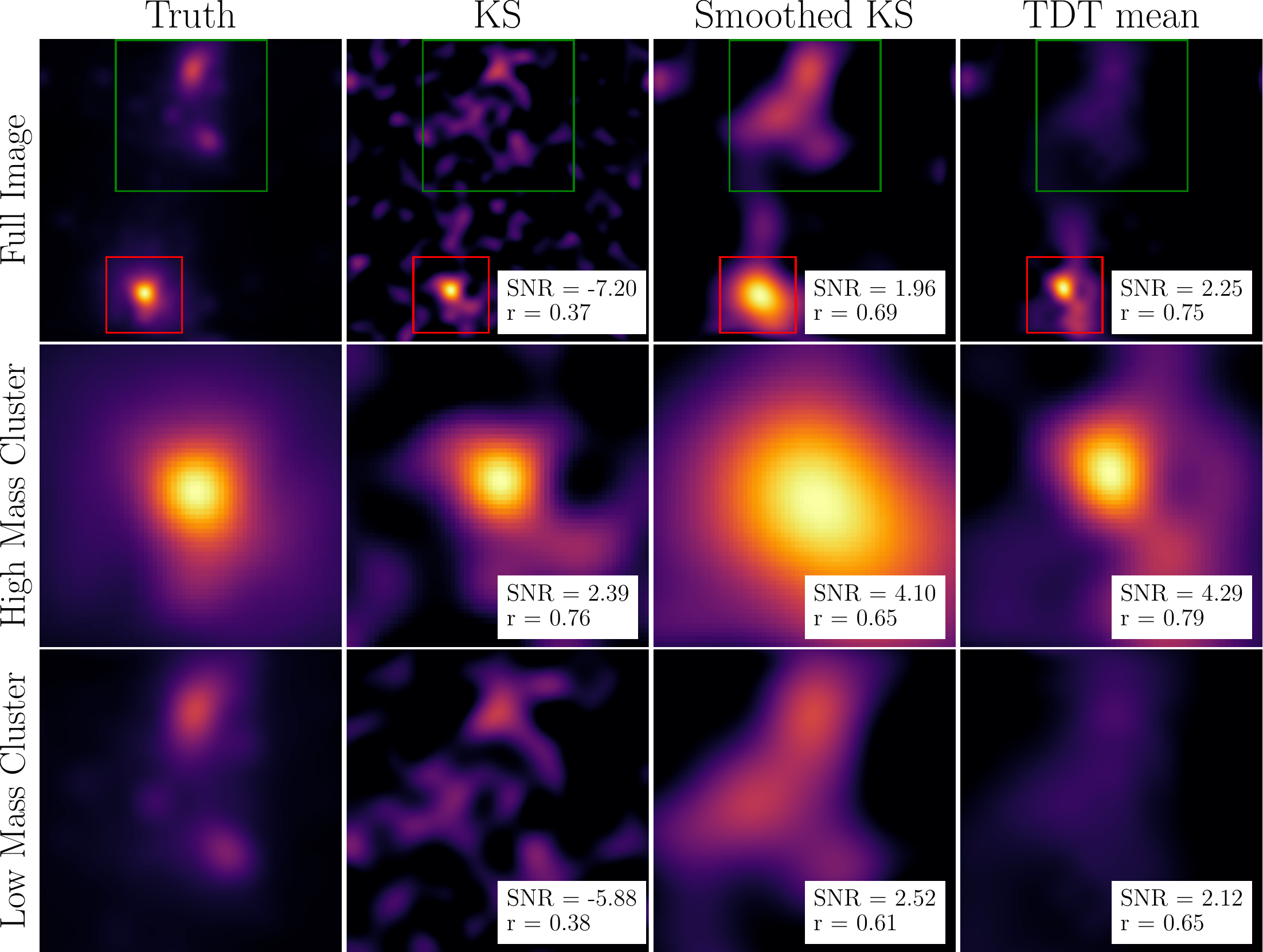}
    \caption[Comparison at high noise on individual clusters]{Comparison of trans-dimensional trees against KS at a noise level of \(n_{gal}=30\), focusing on individual clusters.
    Columns show, from left to right, the ground truth, KS solution, optimally smoothed KS solution and the mean of the trans-dimensional MCMC, respectively.
    The top row shows the full image, corresponding to the top row of \autoref{fig:ks_comparison}.
    The middle and bottom rows zoom in on the cluster with the highest convergence (red box in top row) and the upper low mass cluster (green box in top row), respectively.}
    \label{fig:ks_comparison_clusters}
\end{figure*}

\autoref{fig:ks_comparison_clusters} shows the results for the highest noise level (\(n_{\mathrm{gal}}=30\)), zooming in on the two main clusters.
We have taken a \(8\times8\) pixel box around the high mass cluster at the bottom of the image (red box and middle row in \autoref{fig:ks_comparison_clusters}), and a \(12\times12\) pixel box around the low mass clusters towards to the top (green box and bottom row in \autoref{fig:ks_comparison_clusters}).
The signal-to-noise ratio and correlation coefficients presented have been recalculated within the spatial extent of the respective clusters, and the optimum smoothing for each cluster was also found as described previously for each cluster.
For the cluster with peak convergence, trans-dimensional MCMC again outperforms the optimally smoothed KS method in terms of both overall difference and structural similarities with the ground truth.
For the low mass cluster, the recovery is slightly worse than the optimally smoothed KS in terms of the magnitude of convergence, resulting in a slightly lower SNR\@, although the correlation remains higher.
This shows that our trans-dimensional MCMC is competitive, if not better, than KS with ad-hoc smoothing for a range of cluster masses.

\begin{figure}
    \includegraphics[width=\columnwidth]{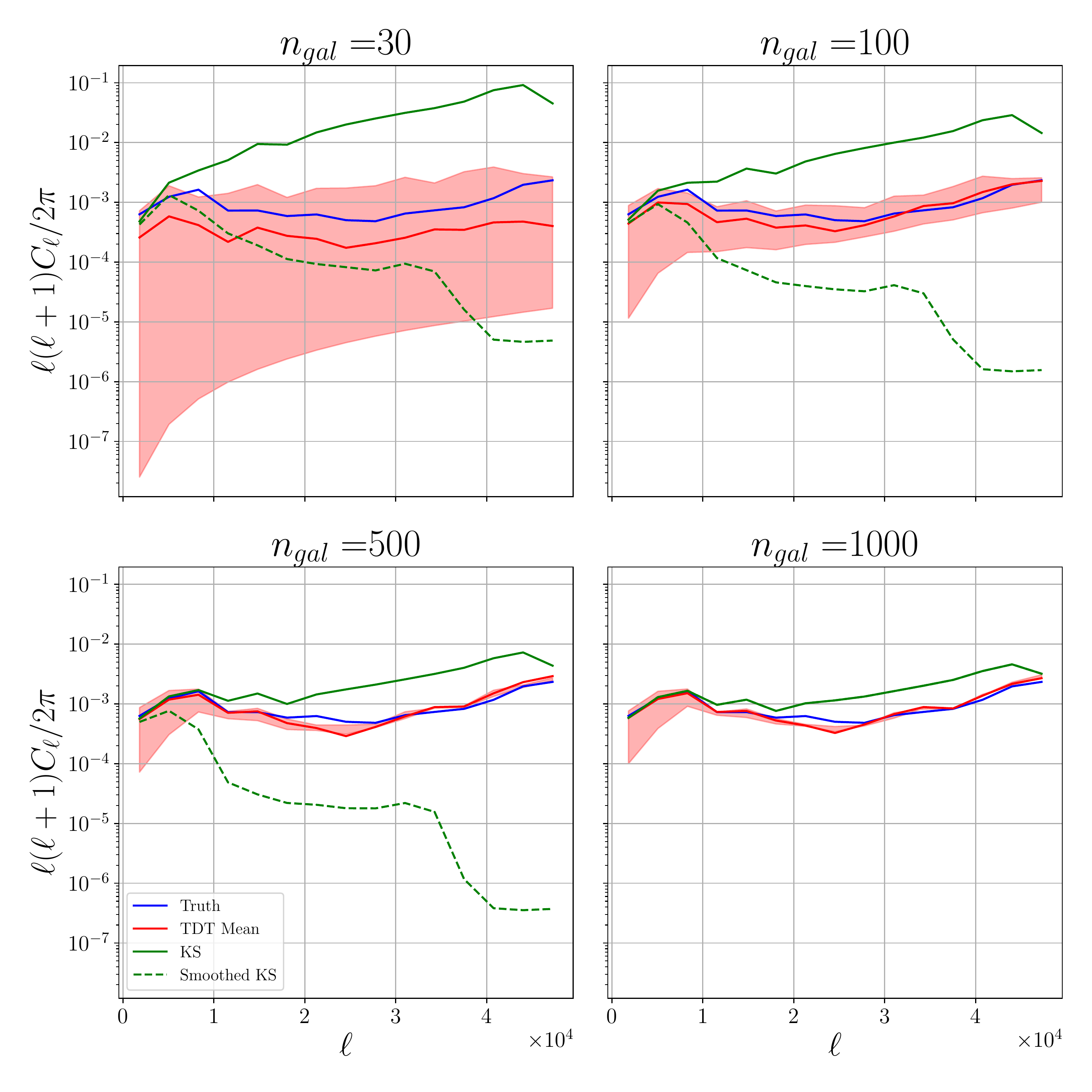}
    \caption[Comparison of recovered convergence spectra]{Power spectra for the KS (solid green), optimally smoothed KS (dashed green) and mean trans-dimensional MCMC (red) solutions (see \autoref{fig:ks_comparison}) compared to the ground truth (blue), at increasing noise levels.
    The light red region shows the 99\% credible interval obtained from trans-dimensional MCMC\@.
    In the lowest noise case (\(n_{\mathrm{gal}}=1000\), bottom right), the optimum smoothing kernel is so small that the spectrum for the smoothed KS solution (dashed green) is indistinguishable from the simple KS solution (solid green).}
    \label{fig:spectra}
\end{figure}

In \autoref{fig:spectra} we show the power spectra for the solutions shown in \autoref{fig:ks_comparison} compared to the ground truth.
At all noise levels, and particularly at the highest \textit{Euclid}-like noise level, the spectrum of the mean trans-dimensional MCMC solution is closer to that of the ground truth than the KS solutions.
Importantly, this is also the case at smaller length scales, where KS is significantly affected by noise and smoothing has the undesirable effect of removing small scale information.
The ground truth spectrum is also shown to be within the uncertainties obtained by MCMC, whereas the KS solutions are inconsistent with the MCMC solution at small length scales.
Uncertainties are seen to be larger at large scales than at small scales.
This is likely a result of the mass-sheet degeneracy, despite there being a tight prior around 0 for the scaling coefficient to alleviate this issue.
These higher uncertainties are thus probably due to the larger scale wavelets which still contribute to the background mean and for which we have wider prior distributions (see \autoref{fig:wavs_ggds}).

\section{Discussion}
\label{sec:discussion}
The trans-dimensional MCMC algorithm is able to significantly decrease the size of the parameter space, searching only a few hundred parameters rather than tens of thousands in the high resolution case (\autoref{fig:bol8}).
The uncertainties also provide further constraints on the lateral extent of the peaks of the convergence field.
All our results also compare favourably to the standard KS method, in terms of the overall fidelity of the reconstruction and in its ability to fully quantify uncertainties (\autoref{fig:ks_comparison}), the recovery of low mass structures (\autoref{fig:ks_comparison_clusters}) and at small length scales (\autoref{fig:spectra}).
Importantly, our results compare favourably to the optimally smoothed KS solution, which is unobtainable in practice.
Furthermore, this sampling problem is far from computationally expensive, thanks to the reduction in parameter space.
All the inversions performed were conducted on a 2020 MacBook Pro with an Apple M1 processor, with the high resolution example taking only 5 hours.

With a new method that more accurately recovers mass maps from noisy shear data and also quantifies full uncertainty, this opens up an exciting prospect of new results relating to dark matter distributions in galaxy clusters.
This method could be applied to high-resolution shear catalogues from the Hubble Frontier Fields \citep{Koekemoer2014} or new data from \emph{Euclid} \citep{Laureijs2011} for new mass-maps and comparisons with, for example, data from the Chandra X-ray Observatory \citep{Weisskopf2000} to investigate the collisional nature of dark matter, as in \citet{Harvey2015}.
\citet{Harvey2015} used a parametric model to reconstruct the density distribution of their colliding galaxy systems \citep{Navarro1997,Jullo2007}.
Our non-parametric method will thus relax some of the physical assumptions made in their density reconstructions, providing additional constraints and validation of their findings.

The main limitations of this trans-dimensional sampler are two-fold.
Firstly, convergence to the peak of the posterior is difficult to determine.
In some cases, the sampler may seem to have converged, then a new tree node is introduced that causes a large jump in posterior probability, bringing the sampler into a new region of parameter space to be explored.
This is a general problem with MCMC methods.
While there is some theoretical understanding about the convergence properties of various probabilistic methods \citep[e.g.][]{Roberts1994,Mengersen1996,Pereyra2016,Durmus2017}, how this translates into number of iterations is not straightforward, and so in practice it is often the case of simply continuing for as long as possible.
A common choice is also to run multiple chains in parallel, starting each chain in a different region of parameter space and converging to the peak of the posterior.
In a standard trans-dimensional MCMC this could be desirable, as one could start with very different models each with a different number of parameters.
For this wavelet tree parameterisation on the other hand, the initial parameter space only has one dimension, by design.
Thus the multiple chains would be starting in very similar points of parameter space, negating the benefit of having parallel chains.
This could be alleviated with some Parallel Tempering techniques \citep{Swendsen1986,Falcioni1999,Brooks2011,Sambridge2013}, whereby each chain samples a slightly modified posterior and there is information exchanged between chains.
For example, each chain will sample the distribution
\begin{equation}
    \pi(\kappa|T) = p(\kappa|\gamma)^{1/T},
\end{equation}
for a given temperature \(T>0\).
For \(T=1\) this is equal to the desired posterior, and for \(T>1\) this is a slightly smoother version of the posterior.
With information being exchanged between chains, there should, in principle, be better mixing of the MCMC samples \citep{Sambridge2013}.

The second main limitation is the handling of very noisy data.
As previously mentioned, when the noise level gets too high the addition or removal of parameters makes little difference to the posterior probability.
Thus any proposal is generally accepted and the tree struggles to grow.
We found this to also be an issue with the smallest wavelet scales, where the addition of a very small wavelet would barely affect the likelihood.
This resulted in anomalously high birth and death acceptance rates among the smallest wavelet scales.
However, there is no tuning parameter (scale and shape of the prior distributions) that has an affect on the acceptance criteria of birth and death proposals (see \hyperref[eqn:birth_acceptance]{Equations~\ref{eqn:birth_acceptance}}~and~\ref{eqn:death_acceptance}).
This is due to the choice of drawing the value of new coefficients from the prior.
While other choices are of course possible, none are physically well-motivated.
As a result it is quite difficult to resolve the smallest and faintest structures, particularly as they get hidden by noisy data.
Nevertheless, we have shown that trans-dimensional MCMC still recovers these smallest scale and lower mass structures better than KS even in high noise settings.

The wavelet tree parameterisation used here is not the only parameterisation that can be used for a trans-dimensional MCMC\@.
Indeed, a popular choice in many geophysical applications is to use Voronoi cells \citep{Voronoi1908,Bodin2012,Young2013b,Dettmer2014,Zhang2018}, where the image space is divided into \(k\) cells defined by a node location such that all the points in a given cell are closer to their cell node than any other cell node.
The value of the field for which one is inverting is generally taken to be constant in each cell.
The nodal density of the cells tends to increase near regions that have the most influence on the posterior, e.g.\ where there is more data or strongly heterogenous regions.
This could conceivably be an excellent option for the mass-mapping problem where the data coverage is dictated by the positions of galaxies on the sky, and gridding may cause holes in the observed shear field.
However, the Voronoi cells may not necessarily form a sparse basis, which may make it difficult to preserve non-Gaussian structures in the resultant convergence map.

\section{Conclusions}
\label{sec:conclusions}
In this work we have used, for the first time, a trans-dimensional MCMC sampler to build cosmological mass-maps in a probabilistic manner, with a sparsity-promoting wavelet prior.
The wavelet tree parameterisation is designed so that the large scale information is resolved first, with smaller scale detail gradually added as required by the data.
Instead of using the standard Laplacian distribution as the sparsity-promoting prior, we exploit the multi-resolution nature of the wavelets and the generalised Gaussian distribution to create a scale-dependent prior that is more sparse for the smaller wavelets.
This approach successfully recovers mass maps from simulated data better than the standard KS method at high noise levels, including levels expected for upcoming surveys, opening up the possibility for new high-resolution mass maps and inferences about the nature of dark matter.

\section*{Acknowledgements}
A.M. is supported by the STFC UCL Centre for Doctoral Training in Data Intensive Science (grant number ST/P006736/1).
A.M.G.F. is grateful for funding from the European Research Council (ERC) under the European Union's Horizon 2020 research and innovation program (grant agreement No 101001601).
The authors are grateful to Rhys Hawkins and Malcolm Sambridge for making their original code publicly available.

\section*{Data Availability}
Code for this work is made available at \url{https://github.com/auggiemarignier/tdtmassmapping}.
It makes use of and adapts the codes made available by Rhys Hawkins at \url{https://github.com/rhyshawkins/TDTbase} and \url{https://github.com/rhyshawkins/TDTWavetomo2D}.
The weak lensing simulation data used is available at \url{http://www.cosmosim.org/}.



\bibliographystyle{mnras}
\bibliography{references}

\begin{thebibliography}{}
\makeatletter
\relax
\def\mn@urlcharsother{\let\do\@makeother \do\$\do\&\do\#\do\^\do\_\do\%\do\~}
\def\mn@doi{\begingroup\mn@urlcharsother \@ifnextchar [ {\mn@doi@}
  {\mn@doi@[]}}
\def\mn@doi@[#1]#2{\def\@tempa{#1}\ifx\@tempa\@empty \href
  {http://dx.doi.org/#2} {doi:#2}\else \href {http://dx.doi.org/#2} {#1}\fi
  \endgroup}
\def\mn@eprint#1#2{\mn@eprint@#1:#2::\@nil}
\def\mn@eprint@arXiv#1{\href {http://arxiv.org/abs/#1} {{\tt arXiv:#1}}}
\def\mn@eprint@dblp#1{\href {http://dblp.uni-trier.de/rec/bibtex/#1.xml}
  {dblp:#1}}
\def\mn@eprint@#1:#2:#3:#4\@nil{\def\@tempa {#1}\def\@tempb {#2}\def\@tempc
  {#3}\ifx \@tempc \@empty \let \@tempc \@tempb \let \@tempb \@tempa \fi \ifx
  \@tempb \@empty \def\@tempb {arXiv}\fi \@ifundefined
  {mn@eprint@\@tempb}{\@tempb:\@tempc}{\expandafter \expandafter \csname
  mn@eprint@\@tempb\endcsname \expandafter{\@tempc}}}

\bibitem[\protect\citeauthoryear{Alsing, Heavens, Jaffe, Kiessling, Wandelt  \&
  Hoffmann}{Alsing et~al.}{2015}]{Alsing2016}
Alsing J.,  Heavens A.,  Jaffe A.~H.,  Kiessling A.,  Wandelt B.,   Hoffmann
  T.,  2015, \mn@doi [Monthly Notices of the Royal Astronomical Society]
  {10.1093/mnras/stv2501}, 455, 4452

\bibitem[\protect\citeauthoryear{Bartelmann \& Schneider}{Bartelmann \&
  Schneider}{2001}]{Bartelmann2001}
Bartelmann M.,  Schneider P.,  2001, \mn@doi [Physics Reports]
  {10.1016/S0370-1573(00)00082-X}, 340, 291

\bibitem[\protect\citeauthoryear{Bodin \& Sambridge}{Bodin \&
  Sambridge}{2009}]{Bodin2009}
Bodin T.,  Sambridge M.,  2009, \mn@doi [Geophysical Journal International]
  {10.1111/j.1365-246X.2009.04226.x}, 178, 1411

\bibitem[\protect\citeauthoryear{Bodin, Sambridge, Gallagher  \&
  Rawlinson}{Bodin et~al.}{2012}]{Bodin2012}
Bodin T.,  Sambridge M.,  Gallagher K.,   Rawlinson N.,  2012, \mn@doi [Journal
  of Geophysical Research: Solid Earth] {10.1029/2011JB008560}, 117

\bibitem[\protect\citeauthoryear{Brooks, Gelman, Jones  \& Meng}{Brooks
  et~al.}{2011}]{Brooks2011}
Brooks S.,  Gelman A.,  Jones G.,   Meng X.-L.,  2011, Handbook of markov chain
  monte carlo.
CRC press

\bibitem[\protect\citeauthoryear{Burdick, Waszek  \& Leki\'{c}}{Burdick
  et~al.}{2019}]{Burdick2019}
Burdick S.,  Waszek L.,   Leki\'{c} V.,  2019, \mn@doi [Earth and Planetary
  Science Letters] {10.1016/j.epsl.2019.115789}, 528, 115789

\bibitem[\protect\citeauthoryear{Cand\`{e}s, Eldar, Needell  \&
  Randall}{Cand\`{e}s et~al.}{2011}]{Candes2011}
Cand\`{e}s E.~J.,  Eldar Y.~C.,  Needell D.,   Randall P.,  2011, \mn@doi
  [Applied and Computational Harmonic Analysis] {10.1016/j.acha.2010.10.002},
  31, 59

\bibitem[\protect\citeauthoryear{Chang et~al.,}{Chang et~al.}{2018}]{Chang2018}
Chang C.,  et~al., 2018, \mn@doi [Monthly Notices of the Royal Astronomical
  Society] {10.1093/mnras/stx3363}, 475, 3165

\bibitem[\protect\citeauthoryear{Clowe, Bradač, Gonzalez, Markevitch, Randall,
  Jones  \& Zaritsky}{Clowe et~al.}{2006}]{Clowe2006}
Clowe D.,  Bradač M.,  Gonzalez A.~H.,  Markevitch M.,  Randall S.~W.,  Jones
  C.,   Zaritsky D.,  2006, \mn@doi [The Astrophysical Journal]
  {10.1086/508162}, 648, L109

\bibitem[\protect\citeauthoryear{Cohen, Daubechies  \& Feauveau}{Cohen
  et~al.}{1992}]{Cohen1992}
Cohen A.,  Daubechies I.,   Feauveau J.-C.,  1992, Communications on pure and
  applied mathematics, 45, 485

\bibitem[\protect\citeauthoryear{Cornish \& Littenberg}{Cornish \&
  Littenberg}{2007}]{Cornish2007}
Cornish N.~J.,  Littenberg T.~B.,  2007, Physical Review D, 76, 083006

\bibitem[\protect\citeauthoryear{Daubechies}{Daubechies}{1992}]{Daubechies1992}
Daubechies I.,  1992, Ten lectures on wavelets.
SIAM

\bibitem[\protect\citeauthoryear{Dettmer, Benavente, Cummins  \&
  Sambridge}{Dettmer et~al.}{2014}]{Dettmer2014}
Dettmer J.,  Benavente R.,  Cummins P.~R.,   Sambridge M.,  2014, \mn@doi
  [Geophysical Journal International] {10.1093/gji/ggu280}, 199, 735

\bibitem[\protect\citeauthoryear{Dodelson}{Dodelson}{2017}]{Dodelson2017}
Dodelson S.,  2017, Gravitational Lensing.
Cambridge University Press, Cambridge, \mn@doi{10.1017/9781316424254}

\bibitem[\protect\citeauthoryear{Donoho}{Donoho}{2006}]{Donoho2006}
Donoho D.,  2006, \mn@doi [IEEE Transactions on Information Theory]
  {10.1109/TIT.2006.871582}, 52, 1289

\bibitem[\protect\citeauthoryear{Durmus, Moulines  \& Saksman}{Durmus
  et~al.}{2017}]{Durmus2017}
Durmus A.,  Moulines E.,   Saksman E.,  2017, \mn@doi [arXiv]
  {10.48550/ARXIV.1705.00166}

\bibitem[\protect\citeauthoryear{Falcioni \& Deem}{Falcioni \&
  Deem}{1999}]{Falcioni1999}
Falcioni M.,  Deem M.~W.,  1999, The Journal of chemical physics, 110, 1754

\bibitem[\protect\citeauthoryear{{Feder} \& {Daylan}}{{Feder} \&
  {Daylan}}{2018}]{Feder2018}
{Feder} R.,  {Daylan} T.,  2018, in American Astronomical Society Meeting
  Abstracts \#231. p. 151.04

\bibitem[\protect\citeauthoryear{Fiedorowicz, Rozo, Boruah, Chang  \&
  Gatti}{Fiedorowicz et~al.}{2022}]{Fiedorowicz2022}
Fiedorowicz P.,  Rozo E.,  Boruah S.~S.,  Chang C.,   Gatti M.,  2022, \mn@doi
  [Monthly Notices of the Royal Astronomical Society] {10.1093/mnras/stac468},
  512, 73

\bibitem[\protect\citeauthoryear{Gallagher}{Gallagher}{2012}]{Gallagher2012}
Gallagher K.,  2012, Journal of Geophysical Research: Solid Earth, 117

\bibitem[\protect\citeauthoryear{Gelman, Carlin, Stern, Dunson, Vehtari  \&
  Rubin}{Gelman et~al.}{2013}]{Gelman2013}
Gelman A.,  Carlin J.~B.,  Stern H.~S.,  Dunson D.~B.,  Vehtari A.,   Rubin
  D.~B.,  2013, Bayesian Data Analysis.
Chapman and Hall/CRC

\bibitem[\protect\citeauthoryear{Geyer \& M{\o}ller}{Geyer \&
  M{\o}ller}{1994}]{Geyer1994}
Geyer C.~J.,  M{\o}ller J.,  1994, Scandinavian journal of statistics, pp
  359--373

\bibitem[\protect\citeauthoryear{Green}{Green}{1995}]{Green1995}
Green P.~J.,  1995, Biometrika, 82, 711

\bibitem[\protect\citeauthoryear{Harvey, Massey, Kitching, Taylor  \&
  Tittley}{Harvey et~al.}{2015}]{Harvey2015}
Harvey D.,  Massey R.,  Kitching T.,  Taylor A.,   Tittley E.,  2015, \mn@doi
  [Science] {10.1126/science.1261381}, 347, 1462

\bibitem[\protect\citeauthoryear{Hawkins \& Sambridge}{Hawkins \&
  Sambridge}{2015}]{Hawkins2015}
Hawkins R.,  Sambridge M.,  2015, \mn@doi [Geophysical Journal International]
  {10.1093/gji/ggv326}, 203, 972

\bibitem[\protect\citeauthoryear{Heavens}{Heavens}{2009}]{Heavens2009}
Heavens A.,  2009, \mn@doi [Nuclear Physics B - Proceedings Supplements]
  {10.1016/j.nuclphysbps.2009.07.005}, 194, 76

\bibitem[\protect\citeauthoryear{Jeffrey et~al.,}{Jeffrey
  et~al.}{2018}]{Jeffrey2018}
Jeffrey N.,  et~al., 2018, \mn@doi [Monthly Notices of the Royal Astronomical
  Society] {10.1093/mnras/sty1252}, 479, 2871

\bibitem[\protect\citeauthoryear{Jeffrey et~al.,}{Jeffrey
  et~al.}{2021}]{Jeffrey2021}
Jeffrey N.,  et~al., 2021, \mn@doi [Monthly Notices of the Royal Astronomical
  Society] {10.1093/mnras/stab1495}, 505, 4626

\bibitem[\protect\citeauthoryear{Jullo, Kneib, Limousin, Eliasdottir, Marshall
  \& Verdugo}{Jullo et~al.}{2007}]{Jullo2007}
Jullo E.,  Kneib J.-P.,  Limousin M.,  Eliasdottir A.,  Marshall P.,   Verdugo
  T.,  2007, New Journal of Physics, 9, 447

\bibitem[\protect\citeauthoryear{Kaiser \& Squires}{Kaiser \&
  Squires}{1993}]{Kaiser1993}
Kaiser N.,  Squires G.,  1993, \mn@doi [The Astrophysical Journal]
  {10.1086/172297}, 404, 441

\bibitem[\protect\citeauthoryear{Karnesis et~al.,}{Karnesis
  et~al.}{2014}]{Karnesis2014}
Karnesis N.,  et~al., 2014, \mn@doi [Physical Review D]
  {10.1103/PhysRevD.89.062001}, 89, 062001

\bibitem[\protect\citeauthoryear{{Klypin}, {Trujillo-Gomez}  \&
  {Primack}}{{Klypin} et~al.}{2011}]{Klypin2011}
{Klypin} A.~A.,  {Trujillo-Gomez} S.,   {Primack} J.,  2011, \mn@doi [The
  Astrophysical Journal] {10.1088/0004-637X/740/2/102}, 740, 102

\bibitem[\protect\citeauthoryear{{Koekemoer} et~al.,}{{Koekemoer}
  et~al.}{2014}]{Koekemoer2014}
{Koekemoer} A.~M.,  et~al., 2014, in American Astronomical Society Meeting
  Abstracts \#223. p. 254.02

\bibitem[\protect\citeauthoryear{Lanusse, Starck, Leonard  \& Pires}{Lanusse
  et~al.}{2016}]{Lanusse2016}
Lanusse F.,  Starck J.~L.,  Leonard A.,   Pires S.,  2016, \mn@doi [Astronomy
  \& Astrophysics] {10.1051/0004-6361/201628278}, 591

\bibitem[\protect\citeauthoryear{Laureijs et~al.,}{Laureijs
  et~al.}{2011}]{Laureijs2011}
Laureijs R.,  et~al., 2011, arXiv preprint arXiv:1110.3193

\bibitem[\protect\citeauthoryear{Malinverno}{Malinverno}{2002}]{Malinverno2002}
Malinverno A.,  2002, Geophysical Journal International, 151, 675

\bibitem[\protect\citeauthoryear{Mallat}{Mallat}{1989}]{Mallat1989}
Mallat S.,  1989, \mn@doi [IEEE Transactions on Pattern Analysis and Machine
  Intelligence] {10.1109/34.192463}, 11, 674

\bibitem[\protect\citeauthoryear{Marignier, McEwen, Ferreira  \&
  Kitching}{Marignier et~al.}{2023}]{Marignier2023}
Marignier A.,  McEwen J.~D.,  Ferreira A. M.~G.,   Kitching T.~D.,  2023,
  \mn@doi [RAS Techniques and Instruments] {10.1093/rasti/rzac010}, 2, 20

\bibitem[\protect\citeauthoryear{McEwen, Feeney, Peiris, Wiaux, Ringeval  \&
  Bouchet}{McEwen et~al.}{2017}]{McEwen2017}
McEwen J.~D.,  Feeney S.~M.,  Peiris H.~V.,  Wiaux Y.,  Ringeval C.,   Bouchet
  F.~R.,  2017, \mn@doi [Monthly Notices of the Royal Astronomical Society]
  {10.1093/mnras/stx2268}, 472, 4081

\bibitem[\protect\citeauthoryear{Mengersen \& Tweedie}{Mengersen \&
  Tweedie}{1996}]{Mengersen1996}
Mengersen K.~L.,  Tweedie R.~L.,  1996, \mn@doi [The Annals of Statistics]
  {10.1214/aos/1033066201}, 24, 101

\bibitem[\protect\citeauthoryear{Minsley}{Minsley}{2011}]{Minsley2011}
Minsley B.~J.,  2011, Geophysical Journal International, 187, 252

\bibitem[\protect\citeauthoryear{Munshi \& Coles}{Munshi \&
  Coles}{2017}]{Munshi2017}
Munshi D.,  Coles P.,  2017, \mn@doi [Journal of Cosmology and Astroparticle
  Physics] {10.1088/1475-7516/2017/02/010}, 2017, 010

\bibitem[\protect\citeauthoryear{Navarro, Frenk  \& White}{Navarro
  et~al.}{1997}]{Navarro1997}
Navarro J.~F.,  Frenk C.~S.,   White S. D.~M.,  1997, The Astrophysical
  Journal, 490, 493

\bibitem[\protect\citeauthoryear{Neal}{Neal}{2012}]{Neal2012}
Neal R.~M.,  2012, in , Handbook of Markov Chain Monte Carlo.
CRC Press, pp 113--162

\bibitem[\protect\citeauthoryear{{Peebles}}{{Peebles}}{1980}]{Peebles1980}
{Peebles} P.~J.~E.,  1980, {The large-scale structure of the universe}

\bibitem[\protect\citeauthoryear{Pereyra}{Pereyra}{2016}]{Pereyra2016}
Pereyra M.,  2016, \mn@doi [Statistics and Computing]
  {10.1007/s11222-015-9567-4}, 26, 745

\bibitem[\protect\citeauthoryear{Pereyra}{Pereyra}{2017}]{Pereyra2017}
Pereyra M.,  2017, \mn@doi [SIAM Journal on Imaging Sciences]
  {10.1137/16m1071249}, 10, 285

\bibitem[\protect\citeauthoryear{{Piana Agostinetti}, Giacomuzzi  \&
  Malinverno}{{Piana Agostinetti} et~al.}{2015}]{Piana2015}
{Piana Agostinetti} N.,  Giacomuzzi G.,   Malinverno A.,  2015, Geophysical
  Journal International, 201, 1598

\bibitem[\protect\citeauthoryear{Porqueres, Heavens, Mortlock  \&
  Lavaux}{Porqueres et~al.}{2021}]{Porqueres2021}
Porqueres N.,  Heavens A.,  Mortlock D.,   Lavaux G.,  2021, \mn@doi [Monthly
  Notices of the Royal Astronomical Society] {10.1093/mnras/stab204}, 502, 3035

\bibitem[\protect\citeauthoryear{Price, Cai, McEwen, Pereyra  \&
  Kitching}{Price et~al.}{2020a}]{Price2020}
Price M.~A.,  Cai X.,  McEwen J.~D.,  Pereyra M.,   Kitching T.~D.,  2020a,
  \mn@doi [Monthly Notices of the Royal Astronomical Society]
  {10.1093/mnras/stz3453}, 492, 394

\bibitem[\protect\citeauthoryear{Price, McEwen, Pratley  \& Kitching}{Price
  et~al.}{2020b}]{Price2020b}
Price M.~A.,  McEwen J.~D.,  Pratley L.,   Kitching T.~D.,  2020b, \mn@doi
  [Monthly Notices of the Royal Astronomical Society] {10.1093/mnras/staa3563},
  500, 5436

\bibitem[\protect\citeauthoryear{Price, McEwen, Cai, Kitching  \& Wallis}{Price
  et~al.}{2021}]{Price2021}
Price M.~A.,  McEwen J.~D.,  Cai X.,  Kitching T.~D.,   Wallis C. G.~R.,  2021,
  \mn@doi [Monthly Notices of the Royal Astronomical Society]
  {10.1093/mnras/stab1983}, 506, 3678

\bibitem[\protect\citeauthoryear{Remy, Lanusse, Jeffrey, Liu, Starck, Osato  \&
  Schrabback}{Remy et~al.}{2022}]{Remy2022}
Remy B.,  Lanusse F.,  Jeffrey N.,  Liu J.,  Starck J.-L.,  Osato K.,
  Schrabback T.,  2022, \mn@doi [arXiv] {10.48550/ARXIV.2201.05561}

\bibitem[\protect\citeauthoryear{Roberts \& Smith}{Roberts \&
  Smith}{1994}]{Roberts1994}
Roberts G.,  Smith A.,  1994, \mn@doi [Stochastic Processes and their
  Applications] {10.1016/0304-4149(94)90134-1}, 49, 207

\bibitem[\protect\citeauthoryear{Sambridge}{Sambridge}{2013}]{Sambridge2013}
Sambridge M.,  2013, \mn@doi [Geophysical Journal International]
  {10.1093/gji/ggt342}, 196, 357

\bibitem[\protect\citeauthoryear{Starck, Themelis, Jeffrey, Peel  \&
  Lanusse}{Starck et~al.}{2021}]{Starck2021}
Starck J.~L.,  Themelis K.~E.,  Jeffrey N.,  Peel A.,   Lanusse F.,  2021,
  \mn@doi [Astronomy \& Astrophysics] {10.1051/0004-6361/202039451}, 649

\bibitem[\protect\citeauthoryear{Swendsen \& Wang}{Swendsen \&
  Wang}{1986}]{Swendsen1986}
Swendsen R.~H.,  Wang J.-S.,  1986, Physical review letters, 57, 2607

\bibitem[\protect\citeauthoryear{Tkal\v{c}i\'{c}, Young, Bodin, Ngo  \&
  Sambridge}{Tkal\v{c}i\'{c} et~al.}{2013}]{Tkalcic2013}
Tkal\v{c}i\'{c} H.,  Young M.,  Bodin T.,  Ngo S.,   Sambridge M.,  2013,
  Nature Geoscience, 6, 497

\bibitem[\protect\citeauthoryear{Van~Waerbeke et~al.,}{Van~Waerbeke
  et~al.}{2013}]{VanWaerbeke2013}
Van~Waerbeke L.,  et~al., 2013, \mn@doi [Monthly Notices of the Royal
  Astronomical Society] {10.1093/mnras/stt971}, 433, 3373

\bibitem[\protect\citeauthoryear{Vorono{\"\i}}{Vorono{\"\i}}{1908}]{Voronoi1908}
Vorono{\"\i} G.,  1908, J. reine angew. Math., 134, 198

\bibitem[\protect\citeauthoryear{Wallis, Price, McEwen, Kitching, Leistedt  \&
  Plouviez}{Wallis et~al.}{2021}]{Wallis2021}
Wallis C. G.~R.,  Price M.~A.,  McEwen J.~D.,  Kitching T.~D.,  Leistedt B.,
  Plouviez A.,  2021, \mn@doi [Monthly Notices of the Royal Astronomical
  Society] {10.1093/mnras/stab3235}

\bibitem[\protect\citeauthoryear{Weisskopf, Tananbaum, Van~Speybroeck  \&
  O'Dell}{Weisskopf et~al.}{2000}]{Weisskopf2000}
Weisskopf M.~C.,  Tananbaum H.~D.,  Van~Speybroeck L.~P.,   O'Dell S.~L.,
  2000, X-Ray Optics, Instruments, and Missions III, 4012, 2

\bibitem[\protect\citeauthoryear{Young, Rawlinson  \& Bodin}{Young
  et~al.}{2013}]{Young2013b}
Young M.~K.,  Rawlinson N.,   Bodin T.,  2013, Geophysics, 78, WB49

\bibitem[\protect\citeauthoryear{Zhang, Curtis, Galetti  \& {de Ridder}}{Zhang
  et~al.}{2018}]{Zhang2018}
Zhang X.,  Curtis A.,  Galetti E.,   {de Ridder} S.,  2018, \mn@doi
  [Geophysical Journal International] {10.1093/gji/ggy362}, 215, 1644

\makeatother
\end{thebibliography}


\end{document}